\documentclass[twocolumn]{revtex4}
\usepackage{graphicx, epsfig}

\def\hmpc{{\, {\rm h}^{-1}~\rm Mpc}}
\def\W{{\cal W}}
\def\PP{{\cal P}}
\newcommand{\fsky}{f_{\rm sky}}

\newcommand{\wj}{\left(
                          \begin{array}{ccc}
                          l_1  &  l_2  & l_3 \\
                            0  &  0    &  0
                          \end{array}
                          \right)}

\newcommand{\wjm}{\left(
                          \begin{array}{ccc}
                          l_1  &  l_2  & l_3 \\
                           m_1  &  m_2   &  m_3
                          \end{array}
                          \right)}

\renewcommand{\thefootnote}{\fnsymbol{footnote}}

\def\cmm2{{\,\rm cm^{-2}}}
\def\cm2{{\,{\rm cm}^2}}
\def\cmm3{{\,{\rm cm}^{-3}}}
\def\gcmm3{{\,{\rm g\,cm^{-3}}}}

\def\fun#1#2{\lower3.6pt\vbox{\baselineskip0pt\lineskip.9pt
  \ialign{$\mathsurround=0pt#1\hfil##\hfil$\crcr#2\crcr\sim\crcr}}}

\begin{document}

\title{Weak Lensing and Dark Energy
\vspace{-0.3cm}}

\author{Dragan Huterer
\vspace{0.1cm}}

\affiliation{\it Department of Physics\\ Enrico Fermi
Institute \\ The University of Chicago \\
Chicago, IL~~60637-1433\vspace{0.15cm}} 

\begin{abstract}
We study the power of upcoming weak lensing surveys to probe
dark energy. Dark energy modifies the distance-redshift relation
as well as the matter power spectrum, both of which affect the
weak lensing convergence power spectrum.  Some dark-energy models
predict additional clustering on very large scales, but this
probably cannot be detected by weak lensing alone due to cosmic
variance. With reasonable prior information on other cosmological
parameters, we find that a survey covering 1000 sq.\ deg.\ down
to a limiting magnitude of $R=27$ can impose constraints comparable to
those expected from upcoming type Ia supernova and number-count
surveys. This result, however, is contingent on the control of
both observational and theoretical systematics. Concentrating on
the latter, we find that the {\it nonlinear} power spectrum of
matter perturbations and the redshift distribution of source
galaxies both need to be determined accurately in order for weak
lensing to achieve its full potential. Finally, we discuss the
sensitivity of the three-point statistics to dark energy.
\end{abstract}

\maketitle

\section{Introduction}

Recent direct evidence for acceleration of the universe
\cite{Riess, Perlmutter} has spurred considerable activity in
finding ways to probe the source of this acceleration, dark
energy \cite{reconstr, Weller, Maor, Goliath} (for a review of
dark energy see Ref.~\cite{Turner_scripta}).  Because dark energy
varies with redshift more slowly than matter, it starts
contributing significantly to the expansion of the universe only
relatively recently, at $z\lesssim 2$. This
component is believed to be smooth (or nearly so), and therefore
detectable mainly through its effect on the expansion rate of the
universe. For these reasons, it is generally believed that type
Ia supernovae (SNe Ia) and number-count surveys of galaxies and
galaxy clusters have the most leverage to probe dark energy, as
they probe the distance and volume in the desired redshift range
\cite{optimal}.  Indeed, planned supernova surveys (e.g.,
SNAP\footnotemark[2]
\footnotetext[2]{http://snap.lbl.gov}) and number-count methods
\cite{Holder, Davis} are expected to
impose tight constraints on the smooth component; for example,
$\sigma (w)\approx 0.05$ from SNAP, assuming a flat universe.

The program of weak gravitational lensing (WL) is primarily
oriented toward mapping the distribution of matter in the
universe. The paths of photons emitted by distant objects and
travelling toward us are perturbed due to the intervening
mass. The weak lensing regime corresponds to the intervening
surface density of matter being much smaller than some critical
value; in that case the observed objects (e.g. galaxies) are
slightly distorted. The weak lensing distortions are small
(roughly at the 1\% level) and one needs a large sample of
foreground galaxies in order to separate the lensing effect from
the ``noise'' represented by random orientations of galaxies.
Therefore, observations of lensed galaxies provides information
on the matter distribution in the universe, as well as the growth
of density perturbations.  Although the potential of WL has been
recognized for around two decades (e.g.\ \cite{early_WL}), only
in the 1990s was there a surge of interest in this area
\cite{Jordi, Blanford, Kaiser_92, Kaiser_98, Jain_Seljak,
Kamionkowski} A unique property of WL is that it is sensitive
directly to the amount of mass in the universe, avoiding the
thorny issue of galaxy-to-mass bias. By measuring ellipticities
of a large number of galaxies, one can in principle directly
reconstruct the mass density field of an intervening massive
object \cite{Kaiser_Squires}. Indeed, the mass reconstruction
of galaxy clusters has been successfully performed on a number of
clusters (for a review, see Ref.~\cite{Mellier}).

An exciting recent development, relevant to this work, was the
discovery of weak lensing by large-scale structure, announced by
four groups \cite{Wittman, Bacon, vW, Kaiser_00}.  The results are in
mutual agreement and consistent with theoretical expectations,
which is remarkable given that they were obtained 
independently. Although current data impose weak constraints on
cosmology (e.g., rule out the Einstein-deSitter Universe with
$\Omega_M=1$), future surveys with larger sky coverage and
improved systematics are expected to impose interesting
constraints on cosmological parameters.

The goal of this work is to assess the power of weak lensing to
constrain dark energy. This analysis therefore complements that
of Huterer \& Turner \cite{optimal}, where the efficacy of SNe Ia and
number-count surveys was considered. We follow the standard
practice of considering dark energy to be a smooth component
parameterized by its energy density (scaled to critical)
$\Omega_X$ and equation-of-state ratio $w=p/\rho$
\cite{Turner_White}.  Dark energy modifies the WL observables by
altering the distance-redshift relation and the growth of density
perturbations. As discussed in Sec.~\ref{sec:NLPS}, the nonlinear
evolution of perturbations also depends on dark energy; this
dependence is much more difficult to calculate and needs to be
calibrated from N-body simulations.  Overall, the dependence of
WL on dark energy is somewhat indirect and expected to be weak,
especially when degeneracy with other cosmological parameters is
taken into account. Nevertheless, we shall show that, provided
systematic errors are controlled and theoretical predictions
sharpened, WL surveys can be efficient probes of dark
energy, comparable to SNe Ia and number-counts. Proposed deep
wide-field surveys such as LSST\footnotemark[2]
\footnotetext[2]{http://dmtelescope.org},
the aforementioned SNAP, and
VISTA\footnotemark[3]
\footnotetext[3]{http://www.vista.ac.uk} will attempt to constrain
dark energy through their WL programs, making our analysis
particularly timely.

Previous work on parameter determination from WL centered mostly
on $\Omega_M$ and $\sigma_8$, the rms density fluctuation in
spheres of $8\hmpc$ \cite{Bernardeau, Jain_Seljak}; here
$H_0=100\,h\,{\rm km/s/Mpc}$ is the Hubble parameter today. Hu
and Tegmark \cite{Hu_Tegmark}, however, used the Fisher matrix
formalism to account for all 8 parameters upon which WL depends,
and assumed dark energy to be the vacuum energy (therefore, fixed
$w=-1$).  We use the same set of parameters, with two changes: we
add $w$, and, guided by the ever-stronger evidence from the
cosmic microwave background (e.g., \cite{boom, DASI, max, Wang}),
we assume a flat universe. To assess the accuracy of parameter
determination, we too use the Fisher matrix machinery, which has
proven to be an extremely efficient and accurate way to forecast
errors in experiments where observables depend on many
parameters.

This paper is organized as follows. In Sec.~\ref{sec:prelim} we
go over the basic formalism and define the notation. In Sections
\ref{sec:pk_l} and \ref{sec:depend} we concentrate on the
convergence power spectrum, and discuss its dependence on
dark energy. Section~\ref{sec:constr} discusses the power of
weak lensing surveys to probe dark energy, while
Sec.~\ref{sec:sys} addresses systematic errors that can lead to
biases in parameter estimation. In Sec.~\ref{sec:3pt} we discuss
the dependence of three-point statistics --- bispectrum and
skewness of the convergence --- on dark energy. We conclude
in Sec.~\ref{sec:concl}.

\section{Preliminaries}\label{sec:prelim}

In this Section we cover the basic formalism of weak
gravitational lensing (for detailed reviews, see
Refs.~\cite{Bartelmann, Mellier}).  We work in the Newtonian
Gauge, where the perturbed Friedmann-Robertson-Walker metric
reads

\begin{eqnarray}
ds^2 &=& -\left(1+2\Phi\right ) dt^2 + 
 a^2(t)\left (1-2\Phi\right ) \times \nonumber \\[0.1cm] 
&& \left [d\chi^2 + r^2(d\theta^2+\sin^2\theta d\phi^2) \right  ]
\end{eqnarray}

\noindent where we have set $c=1$, $\chi$ is the radial  distance,
$\Phi$ is  the gravitational potential, and $k=1$, $0$, $-1$ for closed,
flat and open geometry respectively. We also use the coordinate distance
$r$ which is defined as 

\begin{equation}
r(\chi)=\left \{ 
\begin{array}{cl} 
(-K)^{-1/2}\sinh[(-K)^{1/2}\chi], & \;\mbox{if}\;\;\Omega_{\rm TOT}<1;\\[0.1cm]
{\chi}      ,                     & \;\mbox{if}\;\;\Omega_{\rm TOT}=1;\\[0.1cm]
K^{-1/2}\sin(K^{1/2}\chi) ,       & \;\mbox{if}\;\;\Omega_{\rm TOT} >  1.
\end{array} 
\right.
\end{equation}

\noindent where $K$ is the curvature, $\Omega_{\rm TOT}$ is the
total energy density relative to critical, and $K=(\Omega_{\rm
TOT}-1)H_0^2$.
 
Gravitational lensing produces distortions of images of background
galaxies. These distortions can be described as mapping between the
source plane ($S$) and image plane ($I$) \cite{Hu_White}

\begin{equation}
\delta x_i^S=A_{ij} \delta x_j^I
\end{equation}

\noindent where $\delta {\bf x}$ are the displacement vectors in the
two planes and $A$ is the distortion matrix

\begin{equation}
A=
\left ( 
\begin{array}{cc}
	1-\kappa-\gamma_1	&	-\gamma_2   \\
	-\gamma_2		& 1-\kappa+\gamma_1 
\end{array}
 \right ).  
\end{equation}

\begin{figure}[htbp]
\begin{center}
\epsfig{file=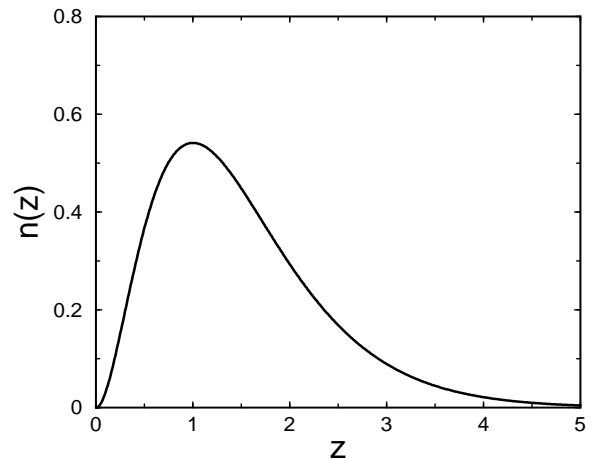, height=2.5in, width=3in}
\caption{The assumed source galaxy distribution $n(z)$.}
\label{fig:prob_tys}
\end{center}
\end{figure}

The deformation is described by the convergence $\kappa$ and
complex shear $(\gamma_1, \gamma_2)$. We are interested in the
weak lensing limit, where $|\kappa|$, $|\gamma|\ll 1$.  The convergence in
any particular direction on the sky ${\bf\hat n}$ is
given by the integral along the line-of-sight

\begin{equation}
\kappa({\bf \hat n}, \chi)=\int_0^{\chi} W(\chi')\, 
\delta(\chi')\, d\chi'
\label{eq:conv}
\end{equation}

\noindent where $\delta$ is the relative perturbation in matter
energy density and

\begin{equation}
W(\chi) = {3\over 2}\,\Omega_M\, H_0^2\,g(\chi)\, (1+z)
\end{equation}

\noindent is referred to as the weight function. Furthermore

\begin{eqnarray}
g(\chi) &=& r(\chi)\int_{\chi}^{\infty} 
d\chi' n(\chi') {r(\chi'-\chi) \over r(\chi')}\\[0.1cm]
&\longrightarrow& {r(\chi) r(\chi_s-\chi) \over r(\chi_s)}
\end{eqnarray}

\noindent where $n(\chi)$ is the distribution of source galaxies in
redshift (normalized so that $\int dz\, n(z)=1$) and the second
line holds only if all sources are at a single redshift $z_s$. We use
the distribution \cite{Wittman}

\begin{equation}
n(z)={z^2\over 2 z_0^3}\, e^{-z/z_0}
\label{eq:prob_tys}
\end{equation}

\noindent with $z_0=0.5$, which peaks at $2z_0=1$ and is shown in
Fig.~\ref{fig:prob_tys}.  Our results depend very weakly on the
shape of the distribution of source galaxies (assuming this distribution
is known, of course).  In particular, if {\it all} source
galaxies are assumed to be at $z=1$, the parameter uncertainties
change by at most $\sim 30\%$ percent. Similarly, the
distribution given by Eq.~(\ref{eq:prob_tys}) which peaks at $z=1.5$
would improve the parameter constraints by 20\% or less. 

Some clarification is needed regarding observability vs.\
theoretical computability of WL quantities.  The quantity that is
most easily determined from observations is shear, which is
directly related to the ellipticities of observed galaxies (in
the weak lensing limit, shear is equal to the average
ellipticity). Shear is given by \cite{Kaiser_98}

\begin{equation}
\gamma_1+i\gamma_2=\frac{1}{2}\left (\psi_{,11}-\psi_{,22}\right
)+i\psi_{,12}	
\end{equation}

\noindent where $\psi$ is the projected Newtonian potential,
$\psi_{,ij}=-2\int g(\chi)\,\Phi_{,ij}\,d\chi$, and commas denote
derivatives with respect to directions perpendicular to the line
of sight. Unfortunately, this quantity is not easily related to
the distribution of matter in the universe and the cosmological
parameters. Convergence, on the other hand, is given by

\begin{equation}
\kappa=\frac{1}{2}\left (\psi_{,11}+\psi_{,22} \right )
\end{equation}

\noindent which (in Limber's approximation) can be directly
related to the distribution of matter through the Poisson
equation (see Eq.~(\ref{eq:conv})), and is convenient for
comparison with theory.  However, it is very difficult
to measure the convergence itself, as convergence depends on the
magnification of galaxies which would somehow need to be measured
(although there may be ways to do this; see Ref.~\cite{Broadhurst})
\footnotemark[2]
\footnotetext[2]{There are two competing effects due to magnification of
galaxies: 1) ``Magnification bias'', the increase in the observed
number of galaxies due to the fact that fainter ones can now be
observed, and 2) increase in the apparent observed area on the
sky due to lensing, which decreases the observed number density
of galaxies.}. Note also that computing the convergence from the
measured shear is difficult in general, since the inversion
kernel is broad and requires knowledge of shear everywhere
\cite{Kaiser_Squires}. In the weak lensing limit,
however, the problem is much easier, since the two-point
correlation functions of shear and convergence are identical.

In this work we use power spectrum of the convergence (defined in
Eq.~(\ref{eq:pk_l}) below) as the principal observable that will
convey information from weak lensing.

\section{Convergence Power Spectrum}\label{sec:pk_l}

\begin{figure}[htbp]
\begin{center}
\epsfig{file=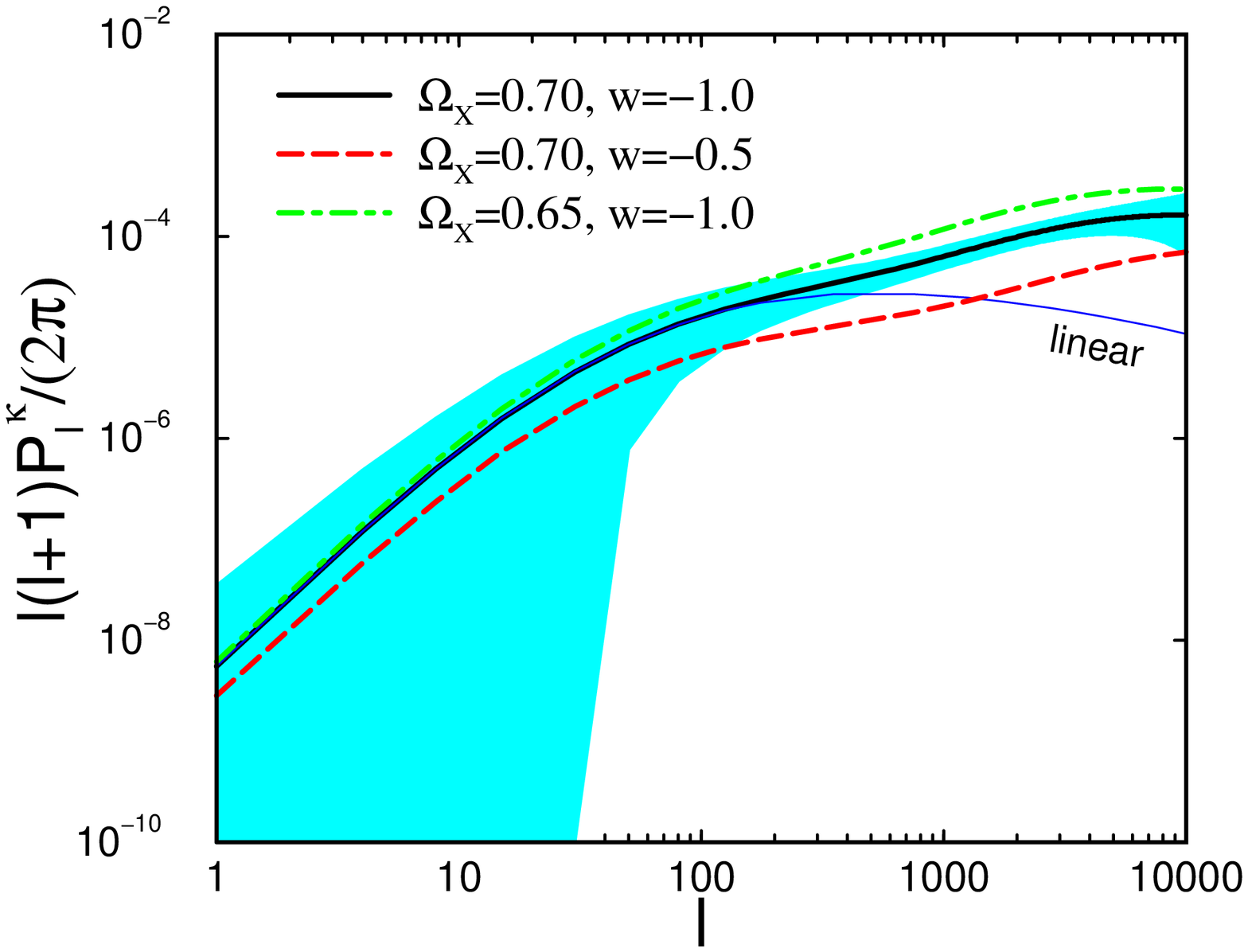, height=2.5in, width=3.0in}\\ 
\epsfig{file=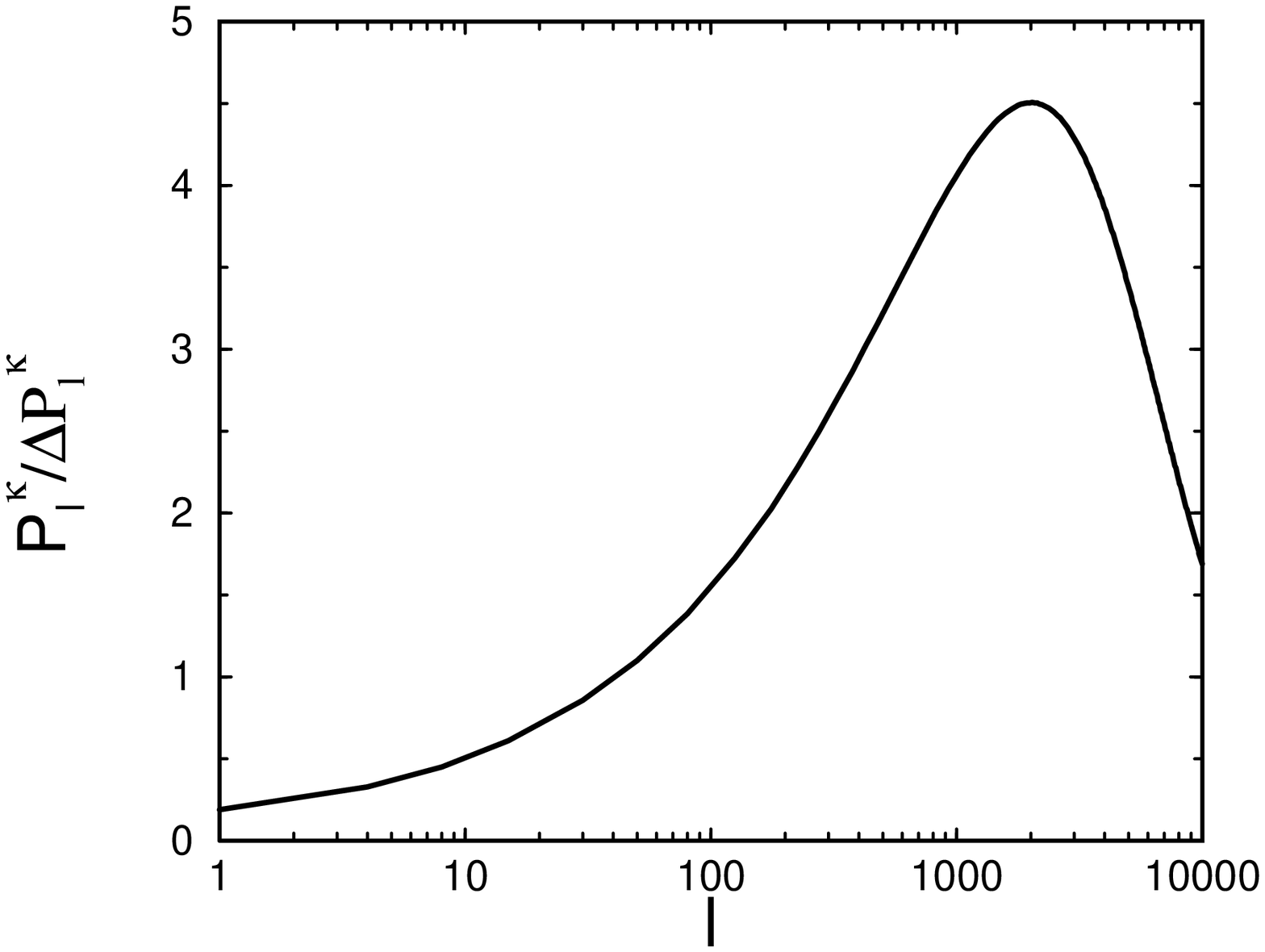, height=2in, width=3.0in}
\caption{{\it Top panel:} The convergence power spectrum for
three pairs of ($\Omega_X$, $w$). The shaded region represents
1-$\sigma$ uncertainties (corresponding to $\Omega_X=0.7$,
$w=-1$ curve) plotted at each $l$. The uncertainties at low $l$ are
dominated by cosmic variance, and those at high $l$ by Poisson
(shot) noise; see Eq.~(\ref{eq:Delta_P}). We also show the
contribution to $P_l^{\kappa}$ from the {\it linear} matter power
spectrum only. {\it Bottom Panel:} $P_l^{\kappa}/\Delta
P_l^{\kappa}$ (``signal-to-noise'') for the convergence power
spectrum for each individual $l$.}
\label{fig:pk_l}
\end{center}
\end{figure}

The convergence can be transformed into multipole space
(see e.g.\ Ref.~\cite{Bartelmann})

$$
\kappa_{lm}=\int d {\bf\hat n} \,\kappa({\bf \hat n}, \chi)\, 
Y_{lm}^*.
$$

The power spectrum of the convergence $P_l^{\kappa}$ is then defined by

$$
\langle \kappa_{lm}\kappa_{l'm'}\rangle =
\delta_{l_1 l_2}\, \delta_{m_1 m_2}  \,P_l^{\kappa}.
$$

Using Limber's approximation --- the fact that the weight
function $W$ is much broader than the physical scale on which the
perturbation $\delta$ varies --- the convergence power spectrum
can be written as

\begin{eqnarray}
P_l^{\kappa} 
&=& \int_0^{\chi_s} d\chi \frac{W^2(\chi)}{r^2(\chi)}
P(l/r(\chi), z)\\[0.1cm]
&=& { 2\pi^2\over l^3}
\int_0^{z_s} dz \,{W^2(z) \,r(z)\over H(z)} \Delta^2(l/r(z), z)
\label{eq:pk_l}
\end{eqnarray}

\noindent where in the second line we assume a flat universe
where $d\chi=dr$. Here $P(k, z)$ is the matter power
spectrum as a function of redshift $z$, and

\begin{equation}
\Delta^2(k, z)={k^3 P(k, z)\over 2\pi^2}
\end{equation}

\noindent is power per unit logarithmic interval in wavenumber,
which we also refer to as the matter power spectrum. 

Power spectrum of the convergence is displayed in the top panel
of Fig.~\ref{fig:pk_l} for three values of ($\Omega_X$, $w$) and
down to scales of about one arcminute ($l=10000$). The
uncertainty in the observed weak lensing spectrum is given by
\cite{Kaiser_92, Kaiser_98}\footnotemark[2]
\footnotetext[2]{Strictly speaking, Eq,~\ref{eq:Delta_P} holds for
Gaussian convergence field only. However, the  non-Gaussianity of the
convergence is milder than that of the matter due to the radial
projection which makes this a good approximation, see
Sec.~\ref{sec:covar}.}

\begin{equation}
\Delta P_l^{\kappa}=
\sqrt{2 \over (2\ell +1)\fsky}\left (
P_l^{\kappa} + {\left< \gamma_{\rm int}^2 \right>\over \bar n} \right ) \,,
\label{eq:Delta_P}
\end{equation}

\noindent where $\fsky = \Theta^2 \pi/129600 $ is the fraction of the
sky covered by a survey of dimension $\Theta$ and $\left<\gamma_{\rm
int}^2\right>^{1/2} \approx 0.4$ is the intrinsic ellipticity of
galaxies.  The first term corresponds to cosmic variance which
dominates on large scales, and the second to Poisson noise which
arises due to small number of galaxies on small scales. Bottom panel
of Fig.\ 2 shows the signal-to-noise $P_l^{\kappa}/\Delta
P_l^{\kappa}$. It is apparent that the bulk of cosmological
constraints comes from multipoles between several hundred and several
thousand. Wider and deeper surveys widen the range of scales with high
signal-to-noise.  Note also that the weak lensing power spectrum is
relatively featureless because of the radial projection
(Eq.~(\ref{eq:pk_l})). It can be characterized by amplitude
(normalization), overall tilt, a ``turnover'' at $l\sim 100$ which is
due to the turnover in the matter power spectrum, and a further
increase at $l\sim 1000$ and flattening at $l\sim 10000$ which are due to
the nonlinear clustering of matter.

The systematic error in shear measurements ideally needs to be
small enough so as not to exceed the statistical error shown in
Fig.~\ref{fig:pk_l}. The maximum allowed systematic error can be
estimated using the following argument\footnotemark[2]
\footnotetext[2]{Proposed by M. Turner.}. The shear variance in
circular aperture of opening angle $\theta$ can be written in
terms of the convergence power spectrum as \cite{Bartelmann}

\begin{eqnarray}
\langle \gamma^2(\theta)\rangle 
&=& 2\pi\int_0^\infty dl\, l P_\kappa^l 
      \left ({J_1(l\theta)\over \pi l\theta}\right )^2 \\
&\approx&   (2\pi)^2 \int_0^\infty {dl\over l} \PP_\kappa^l 
      \left ({J_1(l\theta)\over \pi l\theta}\right )^2 \\
&\simeq & \PP_\kappa^{l=1/\theta} \label{eq:approx}
\end{eqnarray}

\noindent which is the power in unit logarithmic interval
evaluated at $l=1/\theta$ (here
$\PP_\kappa^l=l(l+1)/(2\pi)P_\kappa^l$). The tightest requirement
is on scales of 1 arcmin ($l\approx 2000$), where the fractional
uncertainty in power per unit logarithmic interval is about
$1/200$. Therefore the RMS of shear on scale $\theta$ is given
by

\begin{eqnarray}
\delta \sqrt{\langle \gamma^2(\theta)\rangle} 
&\approx& \delta \sqrt{\PP_\kappa^{l=1/\theta}} \\
&=& {1\over 2} \sqrt{\PP_\kappa^{l=1/\theta}} 
    \left (\delta \PP_\kappa^{l=1/\theta} \over
    \PP_\kappa^{l=1/\theta}\right )\\
&\simeq& {1\over 2}\times \sqrt{10^{-4}}\times {1\over 200}\\
&\approx& 2.5\times 10^{-5} \\
&\lesssim& 0.01\sqrt{\langle \gamma^2(\theta)\rangle}.
\end{eqnarray}

Therefore, the systematic error in individual shear measurements
should be less than $1\%$ in order to be subdominant to statistical
error --- a very challenging requirement indeed!

\section{Dependence on dark energy}\label{sec:depend}

The sensitivity of the convergence power spectrum to dark energy
can be divided into two parts. Dark energy  

a) modifies the background evolution of the universe, and consequently
the geometric factor $W^2(z) r(z)/H(z)$, and

b) modifies the matter power spectrum. 

\noindent We now discuss each of these dependencies.

\subsection{The lensing weight function}\label{sec:weight}

\begin{figure}[htbp]
\begin{center}
\epsfig{file=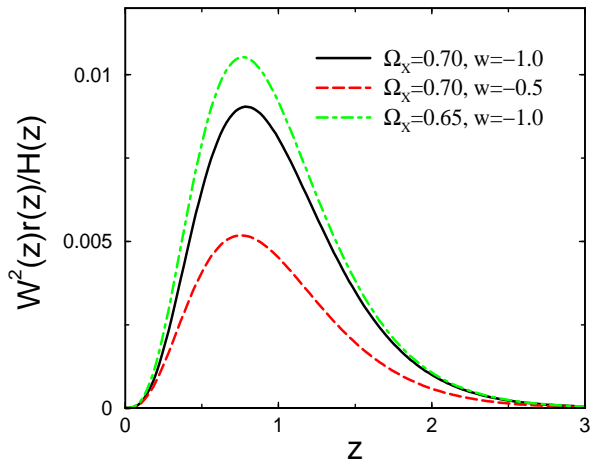, height=2.5in, width=3in}
\caption{The weight function $W^2(z)r(z)/H(z)$ for three pairs of
($\Omega_X$, $w$). }
\label{fig:weight}
\end{center}
\end{figure}

Function $W(z)$ is bell-shaped, and has a maximum at $z\approx
z_s/2$, where $z_s$ is redshift of lensed galaxies, indicating
that lensing is the most effective at distances halfway between the
source and the observer. Since $r(z)$ and $H(z)$ are varying with
redshift monotonically and slowly, the function
$W^2(z) r(z)/H(z)$ will also be bell-shaped with maximum at
$z\gtrsim z_s/2$. $W(z)$, $r(z)$ and $1/H(z)$ all decrease with
increasing $w$, and therefore the total weight decreases. As
$\Omega_X$ decreases, $r(z)$ and $1/H(z)$ decrease but $W(z)$
increases, and the latter prevails; see
Fig.~\ref{fig:weight}. Therefore, changing $w$ ($\Omega_X$) makes the
normalization and total weight change with the same (opposite)
sign, leading to large (small) change in $P_l^{\kappa}$ on large
scales; see Fig.~\ref{fig:pk_l}.


\subsection{The Matter Power Spectrum}\label{sec:mat_ps}

The matter power spectrum can be written as

\begin{eqnarray}
\Delta^2(k, z)
&=&\delta_H^2\left (k\over H_0\right )^{3+n}
\!\! T^2(k, z)\, {D^2(z)\over D^2(0)} \, T_{NL}(k, z)
\end{eqnarray}

\noindent where $\delta_H$ is perturbation on Hubble scale today,
$T(k)$ is the transfer function, $D(z)/D(0)$ is the growth of
perturbations in linear theory relative to today, and $T_{NL}(k,
z)$ is the prescription for nonlinear evolution of the power
spectrum.

In the presence of dark energy, the matter power spectrum will be
modified as follows. 

$\bullet$ The normalization $\delta_H$ increases with increasing
$\Omega_X$ and decreasing $w$. This happens because the growth of
structure is suppressed in the presence of dark energy, and the
observed structure today can only be explained by a larger
initial amount of perturbation. We choose to normalize the
results to COBE measurements \cite{Bennett}, and adopt the fit to
COBE data of Ma et al.\ (\cite{Ma_QCDM}, heretofore Ma QCDM)

\begin{eqnarray}
\delta_H&=&2\times 10^{-5} \alpha(0)^{-1} 
\Omega_M^{\,c_1+c_2\ln(\Omega_M)}\nonumber\\[0.1cm]
&&\times \exp\left [ c_3(n-1)+c_4(n-1)^2\right ]
\end{eqnarray}
 
\noindent where $c_{1\ldots 4}$ and $\alpha_0$ are functions of
$\Omega_X$ and $w$ and are given in Ma QCDM.  Since the COBE
normalization for $\Lambda$CDM models is accurate to between 7\%
and 9\% \cite{Bunn_White, Liddle_Lyth}, we adopt the
accuracy of 10\% for the dark-energy case.

$\bullet$ The transfer function for cosmological models with
neutrinos and the cosmological constant is given by fits of Hu
and Eisenstein \cite{Hu_transfer}, which we adopt in our
analysis. These formulae are accurate to a few percent for the
currently favored cosmology with low baryon abundance. Dark
energy will not directly modify the transfer function, except
possibly on the largest observable scales (of size $\sim
H_0^{-1}$), where dark energy may cluster. This signature can be
ignored, as it shows up on scales too large to be probed by WL;
we further discuss this in Sec.~\ref{QCDM_clustering}.

$\bullet$ The linear theory growth function $D(z)=
\delta(z)/\delta(0)$ can be computed from the fitting formula for
the dark-energy models given in Ma QCDM, which generalizes the
$\Lambda$CDM growth function formula of Carroll et al.\
\cite{Carroll}. We use this fitting function, noting that its
high accuracy ($\leq 2\%$) justifies avoiding the alternative of
repeatedly evaluating the exact expression for the growth
function (e.g.\ \cite{Peebles}, p.\ 341).

$\bullet$ The last, and most uncertain, piece of the puzzle is
the prescription for the non-linear evolution of density
perturbations. This is given by the recipe of Hamilton et al.\
\cite{Hamilton} as implemented by Peacock and Dodds (\cite{PD};
heretofore PD), as well as Ma (\cite{Ma_LCDM}, heretofore Ma
$\Lambda$CDM).  These two prescriptions were calibrated for
$\Lambda$CDM models, although the PD formula seems to adequately
fit models with $w>-1$ (M. White, private communication). Ma QCDM
prescription \cite{Ma_QCDM}, on the other hand, gives explicit
formulae for the nonlinear power spectrum in the presence of dark
energy (i.e., a component with $w\geq -1$).  Unfortunately, we
found that the PD and Ma QCDM prescriptions agree (to $\sim
15$\%) only at values of $w$ where Ma QCDM was tested. At other
values of $w$ the maximum disagreement between the two is up to
50\%, and it is not clear which fitting function, if any, is to
be used.  We choose to use the PD prescription primarily because
it is implemented for all $w$ and therefore facilitates taking
the derivative with respect to $w$ needed for the Fisher
matrix. In Sec.~\ref{sec:NLPS} we explore the possible parameter
biases due to the uncertain calibration of the nonlinear power
spectrum.

\begin{figure}[htbp]
\begin{center}
\epsfig{file=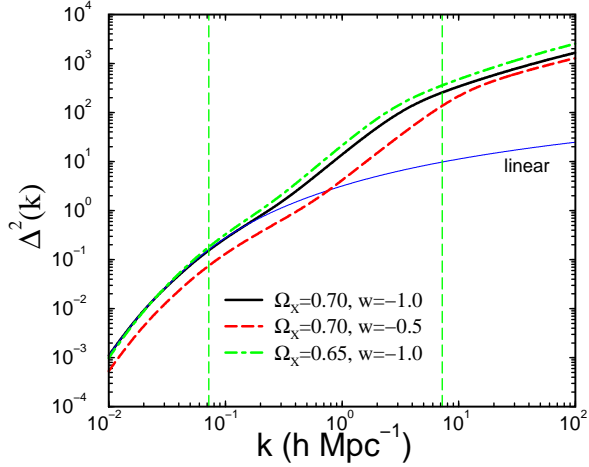, height=2.5in, width=3.0in}
\caption{The matter power spectrum at $z=0$ for three pairs of
$(\Omega_X, w)$. Linear power spectrum corresponding to the
fiducial spectrum is shown by the thin solid curve.  Vertical
lines delimit the interval which contributes significantly to the WL
convergence power spectrum, roughly corresponding to $100\leq l \leq
10000$. It can be seen that the ability to determine cosmological
parameters will depend critically upon the knowledge of the nonlinear power
spectrum.  }
\label{fig:matter_ps}
\end{center}
\end{figure}

Figure~\ref{fig:matter_ps} shows the matter power spectrum at
$z=0$ for three pairs of $(\Omega_M, w)$. When $\Omega_X$ or $w$
are varied, the growth and normalization change affecting all scales
equally. Varying $\Omega_X$ also changes the transfer function
(at $k\gtrsim 0.02\hmpc$). On smaller scales ($k\gtrsim
0.2\hmpc$), the non-linear power spectrum is further affected by
dark energy. 

\subsection{Angular Scale --  Physical Scale Correspondence}

\begin{figure}[htbp]
\begin{center}
\epsfig{file=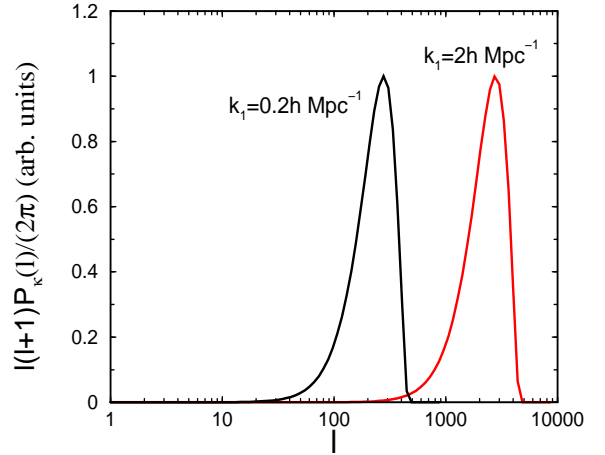, height=2.5in, width=3.0in}
\caption{Power spectrum of the convergence assuming matter power
spectrum is a delta-function at $k_1$,
shown for two different values of $k_1$. This shows the
correspondence between physical and angular scales (for $z_s=1$
and our fiducial $\Lambda$CDM cosmology). }
\label{fig:ps_delta}
\end{center}
\end{figure}

To illustrate the correspondence between wavenumbers $k$ and multipoles $l$, 
let us assume the matter power spectrum peaked at a single multipole $k_1$

\begin{equation}
\Delta^2(k)=\langle \delta^2\rangle k_1\delta(k-k_1)
\end{equation}

\noindent normalized so that $\int\Delta^2(k)d\ln k=\langle
\delta^2\rangle$ (here $\langle \delta^2\rangle$ is the auto-correlation
function of density contrast in real space).  Then, assuming for
simplicity that all sources are at a single redshift $z_s$, we
have

\begin{equation}
l(l+1)P_l^{\kappa}/(2\pi)
  \propto  {l^3\over k_1^4} \left (1-{l\over r(z_s)k_1}\right )^2 
\end{equation}

\noindent for $l< k_1r(z_s)$, and zero for $l\geq k_1r(z_s)$. The
plot of the convergence power spectrum is given in
Fig.~\ref{fig:ps_delta} for two values of $k_1$. The multipole
power peaks at $l=3/5\, k_1 r(z_s)$.  Assuming a survey with
$z_s=1$, the scale at which the non-linear effects become
significant, $k\approx 0.2\hmpc$, corresponds to $l\approx 300$.
Our constraints mostly come from angular scales $l\sim 1000$,
corresponding to $k\sim 1\hmpc$. The bulk of WL constraints
therefore comes from non-linear scales.

\section{Constraints on  Dark Energy}\label{sec:constr}

\subsection{The Fisher matrix formalism }\label{sec:Fisher}

The fact that the relatively featureless $P_l^{\kappa}$ depends
upon a number of cosmological parameters directly leads to
parameter degeneracies and limits the power of weak lensing to
measure these parameters independently of other probes, even for
the case of a full-sky survey.  To estimate how accurately
cosmological parameters can be measured, we use the Fisher matrix
formalism \cite{Fisher}.  This method has already been used to
forecast the expected accuracies from CMB surveys
\cite{Zaldarriaga_Seljak, Eis_Hu_Teg}, SNe Ia \cite{SN_Fisher,
optimal, Goliath} and number counts \cite{Holder_01} and was
found to agree very well with direct Monte-Carlo error
estimation. Its considerable advantage over Monte Carlo is that
it does not require simulations and analyses of data sets, but
only a single evaluation of a simple analytic
expression. Furthermore, the Fisher matrix formalism allows easy
inclusion of Bayesian priors and constraints from other methods.

The Fisher Matrix is defined as the second derivative of the
negative log-likelihood function

\begin{eqnarray}
F_{ij} &=& 
-\left< \partial^2 \ln L \over \partial p_i 
\partial p_j \right>_{\bf x}\\[0.15cm]
&=& \,\sum_{l}\frac{1}{(\Delta P_l^{\kappa})^2}
 {\partial P_l^{\kappa} \over \partial p_i} 
 {\partial P_l^{\kappa} \over \partial p_j}\,.
\label{eq:Fisher}
\end{eqnarray}

\noindent where $L$ is the likelihood of the observed  data set ${\bf
x}$ given the parameters $p_1 \ldots p_n$. The second line
follows by assuming that $L$ is Gaussian in the observable
$P_l^{\kappa}$, which is a good assumption for small departures
around the maximum. 

In practice we do not estimate the power spectrum at every
multipole $l$, but rather bin $P_l^{\kappa}$ in 17 bins. We
explicitly checked that binning makes no significant difference
in our results (this is not surprising, as the convergence power
spectrum doesn't have features that would get washed out by
moderate-resolution binning). We considered $P_l^{\kappa}$ at
$100\leq l\leq 10000$, corresponding to angles between 1
arcminute and 2 degrees on the sky. Variations in the minimum and
maximum $l$ do not change any of our results, as very large and
very small scales are dominated by cosmic variance and Poisson
noise respectively.

Finally, we need to choose steps in parameter directions when
taking numerical derivatives. We choose the steps to be 5\% of
the parameter values, making sure to take two-sided derivatives.


\subsection{The fiducial cosmology and fiducial survey}\label{sec:fiducial}

Finally, we need to choose the fiducial survey, i.e.\ sky
coverage and depth of the survey. We do not consider any single
experiment in particular, but rather adopt numbers roughly
consistent with proposed dedicated wide-field surveys expected to
become operational in several years. We assume a survey covering
1000 sq.\ deg.\ down to a limiting magnitude $R=27$;
dependence of the results upon these two parameters is discussed
in Sec.~\ref{sec:tomo}. Surveys of this power are not yet
operational, but are expected in the near future with results
perhaps by the end of this decade. To convert from magnitudes to
surface density of galaxies, we use the correspondence from
Herschel and Hubble Deep Fields \cite{Metcalfe}, which
for our fiducial numbers implies 165 gal/arcmin$^2$. We assume
that the {\it only} sources of noise are statistical: cosmic
variance which dominates on large scales, and shot-noise dominant
on small scales. We discuss the effect of systematics in
Sec.~\ref{sec:sys}.

\subsection{Parameter space}

Power spectrum of the convergence depends on 7 parameters:
$\Omega_X$, $w$, $\Omega_M h^2$, $\Omega_B h^2$, $\delta_H$, $n$,
and $m_\nu$, where $\Omega_B$ is the energy density in baryons
(relative to critical), $n$ is the spectral index of scalar
perturbations, and $m_\nu$ the neutrino mass summed over all
species. In addition, $P_l^{\kappa}$ depends upon the
redshift distribution of source galaxies. Throughout, we use a
fiducial model that fits well all experiments so far:
$\Omega_X=1-\Omega_M=0.7$ (flat universe assumed), $h=0.65$,
$\Omega_B h^2=0.019$, $n=1.0$, and $\delta_H$ inferred from COBE
measurements as described in Sec.~\ref{sec:mat_ps}. The mass of
neutrino species is quite uncertain, but, according to solar
neutrino experiments, likely to be between zero and a few eV; we
adopt $m_\nu=0.1{\rm eV}$. 

We would like to get an insight in parameter degeneracies, in
particular between the equation of state ratio $w$ and other
parameters. To do that, we compute the correlation between $w$
and other parameters. 
The correlation coefficient is given by

\begin{equation}
\rho(w, p_i)={{\rm Cov}(w, p_i) \over \sqrt{{\rm Cov}(w, w) 
{\rm Cov}(p_i, p_i)}}
\end{equation}

\noindent where ${\rm Cov}(p_i, p_j)=F_{ij}^{-1}$ is an element of the
covariance matrix. Because imposing priors would alter the
covariance matrix and confuse its interpretation, at this point we add
no priors except for COBE normalization (10\% in $ \delta_H$) and
perfect knowledge of galaxy redshifts.

The most significant correlations are $\rho(w, \Omega_X)=-0.96$,
$\rho(w, \Omega_B h^2)=-0.83$, and $\rho(w, m_\nu)=0.81$. 
We find that these and other correlations are typically very
dependent on the fiducial model and the assumed prior.
Finally, we examine the eigenvalues and eigenvectors of the
Fisher matrix. The combination $0.61\,\Omega_X+0.21\,w$ is
determined to an accuracy of about $0.03$; this is the
best-determined combination containing significant components in
$\Omega_X$ and $w$ directions. The most poorly combination of all
is one almost entirely in the $w$-direction:
$0.97\,w-0.21\,\Omega_X$; it is determined to about 0.4.

\subsection{Bayesian Priors}\label{sec:priors}

Without {\it any} prior information on cosmological parameters,
weak lensing imposes very weak constraints on dark energy (and
other parameters as well). The reason is that the power spectrum
of the convergence is featureless, owing to the fact that it
represents the radial projection of the density contrast. Unlike
the CMB spectrum, it lacks bumps and wiggles that would help
break parameter degeneracies.  Constraints rapidly improve,
however, if the redshift distribution of source galaxies is
known. We assume this to be the case; indeed, photometric
redshift techniques already show that distribution of source
galaxies in weak lensing surveys will be determined independently
of cosmological parameters (e.g., \cite{Hogg}). Exact
knowledge of the source distribution is obviously a strong and
perhaps unrealistic assumption, and in Sec.~\ref{sec:sys} we
explore what happens when the uncertainties are included.

There is no reason to expect that any cosmological probe alone
should carry the burden of determining all cosmological
parameters. Indeed, a number of cosmological parameters are
already pinned down quite accurately by other means. In about 10
years, when powerful weak lensing surveys we consider complete
their observational programs, parameters inferred from the CMB
(such as $\Omega_M h^2$, $\Omega_B h^2$ and $n$) will be
determined to an accuracy of several percent
\cite{Eis_Hu_Teg}. The neutrino mass, on the other hand, is
poorly known today, but in the near future it is likely to
be constrained by a combination of CMB, Ly-$\alpha$ forest
\cite{Ly_alpha}, as well as solar and atmospheric neutrino
measurements.

For these reasons, we include Gaussian priors on cosmological
parameters (other than $\Omega_X$ and $w$). We consider two sets
of priors, and call them ``COBE+photo-z'' and ``Planck (T)''. The
former set of priors is a weak one: we only include the 10\%
uncertainty in COBE normalization and, as mentioned above,
knowledge of the distribution of background galaxies.  The latter
set is a moderate one, corresponding to the COBE+photo-z prior,
plus the constraints expected from the Planck mission with
temperature information only (Table 2 of Ref.~\cite{Eis_Hu_Teg}):
$\sigma(\ln \Omega_M h^2)=0.064$, $\sigma(\ln
\Omega_Bh^2)=0.035$, $\sigma(n)=0.04$, and
$\sigma(m_\nu)=0.58$\footnotemark[2]
\footnotetext[2]{Strictly speaking, the correct way
to add the CMB priors would be to add the WL and CMB Fisher
matrices. This procedure would correctly account for breaking of
the WL parameter degeneracies by the CMB. We opt, however, to
just add the priors to the diagonal elements of the WL Fisher
matrix. This effectively assumes other parameters to be
constrained within some limits, regardless of what experiment
those constraints come from.}. We note, however, that details of
the second prior do not change the results much; for example,
using the considerably weaker assumptions corresponding to MAP
mission (with temperature only) instead of Planck (T), errors in
$\Omega_X$ and $w$ degrade by only 10\% and 5\%
respectively. Similarly, using the very strong prior of Planck
constraints (temperature and polarization) combined with those
from Sloan Digital Sky Survey
(SDSS\footnotemark[3]
\footnotetext[3]{http://www.sdss.org}), the constraints improve only by
about 20\%. The reason for this weak dependence on the
prior is easy to understand: by assuming the knowledge of the
distribution of source galaxies and adding other priors, we have
broken the major degeneracy between $\Omega_X$, $w$ and other
parameters; further information on other parameters leads to
small improvements in the constraints on dark energy.

\subsection{Results}\label{sec:results}

\begin{figure}[htbp]
\begin{center}
\epsfig{file=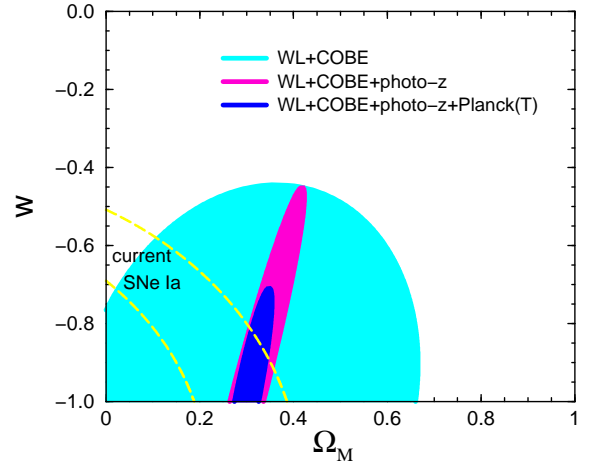, height=2.5in, width=3in}
\caption{68\% C.L.\ constraints on $\Omega_M$ and $w$ for two
values of $m_\nu$. We assume a survey of 1000 sq.\ deg.\ down to
27th magnitude in $R$-band, assume knowledge of the distribution
of source galaxies, and impose the Planck (T) prior on other
parameters \cite{Eis_Hu_Teg}. The strength of the
constraints does not depend sensitively on the set of priors, but
does depend on the fiducial model (e.g., the neutrino mass). For
orientation, current 1-$\sigma$ constraints from 42 type Ia
supernovae \cite{Perlmutter} are also shown.}
\label{fig:constr}
\end{center}
\end{figure}

An example of the constraints that weak lensing can impose on
dark energy is shown in Fig.~\ref{fig:constr}. Here we show the
68\% constraint regions for our fiducial WL survey (1000 sq.\
deg.\ down to 27th mag)  with  several sets of priors on other parameters.
The ellipse is oriented so that increase in $w$ is degenerate
with increase in $\Omega_M$, which is opposite of what we would
expect; this is due to the fact that we assume galaxy redshifts
to be known \footnotemark[4] \footnotetext[4]{In general, priors
on other cosmological parameters will change the orientation of
the constraints in the $\Omega_M$-$w$ plane.}.  Table 1 lists the
uncertainties using two sets of priors. Weak lensing is
potentially a strong probe of dark energy: the 1-$\sigma$
uncertainties in $\Omega_X$ and $w$ are $\sim 5\%$ and 20-40\%
respectively (depending on the set of priors), which is somewhat
weaker than statistical errors expected from future SNe Ia and
number-count surveys. We emphasize that these numbers are the
best ones possible given the survey specifications; systematic
errors may weaken the constraints (see Sec.~\ref{sec:sys}). It is
also true, however, that weak lensing tomography can
significantly {\it improve} these constraints (more on that in
Sec.~\ref{sec:tomo}).

\newpage

\begin{center}
{TABLE 1\\[4pt] \scshape \rule[-2mm]{0mm}{5mm} Constraints on
dark energy}
\nopagebreak
\begin{tabular}{lcc}\hline\hline
      & \multicolumn{2}{c}{\rule[-2mm]{0mm}{6mm}Prior}  \\ \cline{2-3}
\rule[-3mm]{0mm}{8mm}  & COBE + photo-z & Planck (T)\\\hline\\[-0.15cm]

$\sigma(\Omega_X)$
	&   0.08	
	&   0.04 \\[0.1cm]
$\sigma(w)$
	&	0.36
	&	0.19  \\[0.1cm]
\label{tab:results}
\end{tabular}
\end{center}

\begin{figure}[!h]
\begin{center}
\epsfig{file=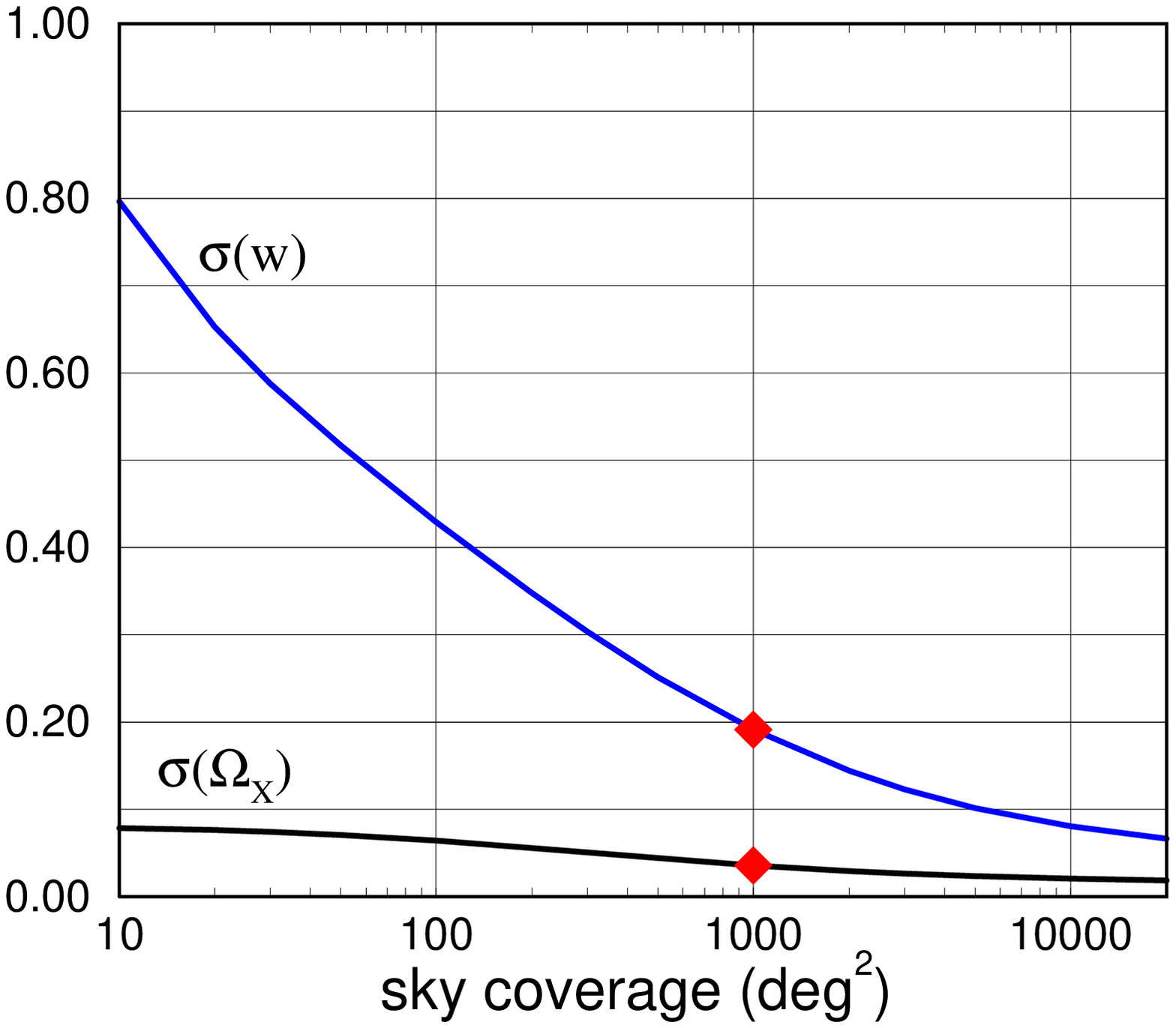, height=2.5in, width=3in}
\epsfig{file=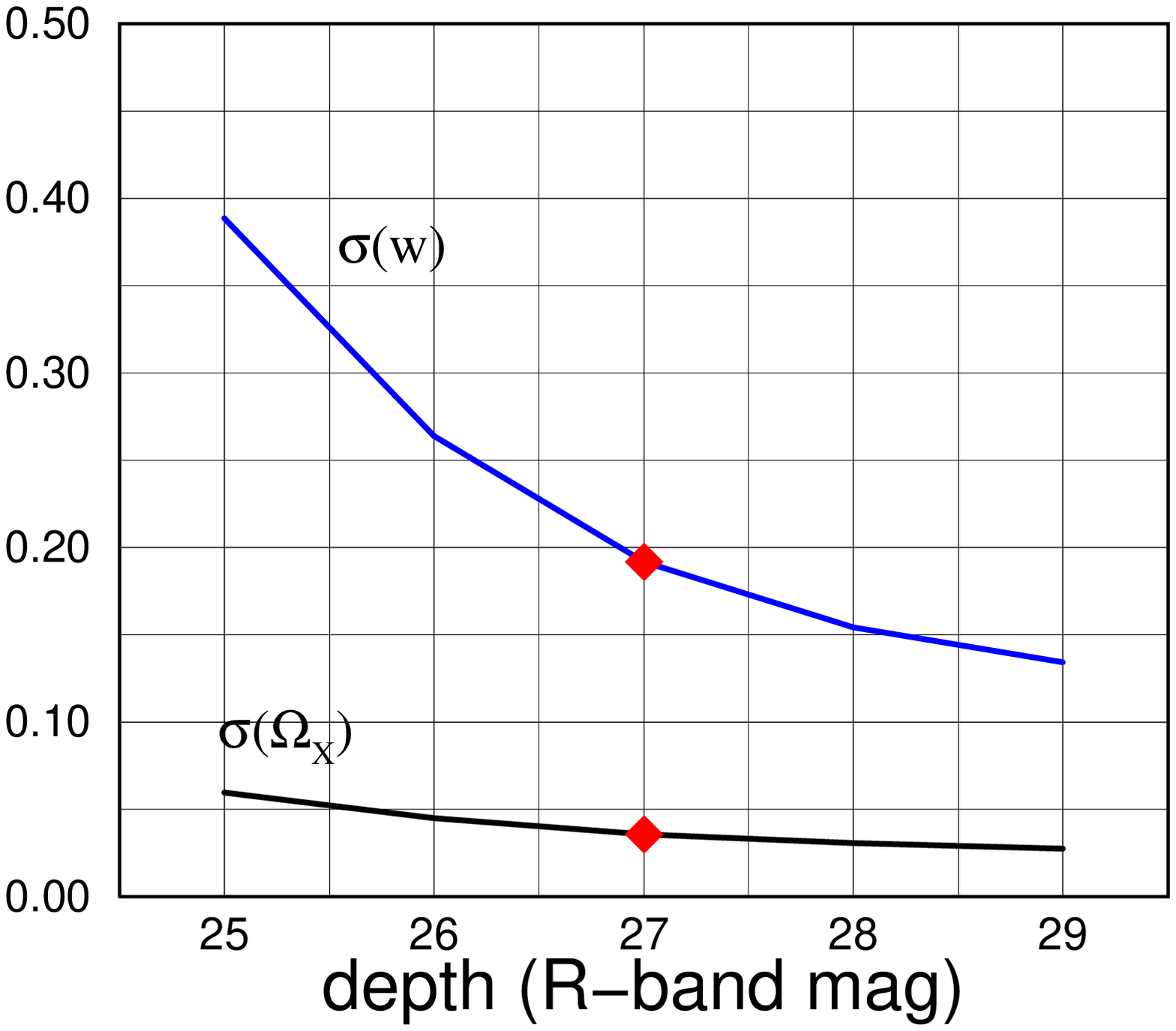, height=2.5in, width=3in}
\caption{ Dependence of $\sigma(\Omega_X)$ and $\sigma(w)$ on the
survey parameters. In each case we assume Planck (T) priors on
other cosmological parameters. Diamonds denote the  fiducial
values. {\it Top panel}: 1-$\sigma$ uncertainties on $\Omega_X$
and $w$ as a function of sky coverage of the survey. We assume a
fixed depth of 27th magnitude in $R$-band. {\it Bottom panel}:
1-$\sigma$ uncertainties on $\Omega_X$ and $w$ as a function of
depth of the survey, assuming a fixed sky coverage of 1000 sq.\
deg.}
\label{fig:sigma_vs_fsky}
\end{center}
\end{figure}

Top panel of Fig.~\ref{fig:sigma_vs_fsky} shows the dependence of
the uncertainties in $\Omega_X$ and $w$ on the sky coverage,
holding the depth of the survey fixed at 27th mag. Lower panel of
the same Figure shows dependence of the uncertainties on the
depth of the survey, holding the sky coverage fixed at 1000 sq.\
deg.  The constraints on $w$ depend quite strongly on the depth
of the survey --- for example, constraint on $w$ would improve by
a factor of two by increasing the coverage of the survey to 5000
sq.\ deg. The dependence on the depth is also significant, but
probably complicated by some practical problems; for example,
galaxy overlap. Therefore, future
surveys with very deep and/or wide sky coverage will be  especially
effective probes of dark energy.

\subsection{Power spectrum Tomography}\label{sec:tomo}

One way to extract more information out of the data would be to
divide the lensed galaxies in several redshift bins and measure
the convergence power spectrum in each bin, as well as the cross
power spectrum between bins. This procedure, the power spectrum
tomography, should be fully feasible with upcoming surveys
because redshifts of source galaxies are going to be known
quite accurately through photometric techniques.

\begin{figure}[htbp]
\begin{center}
\epsfig{file=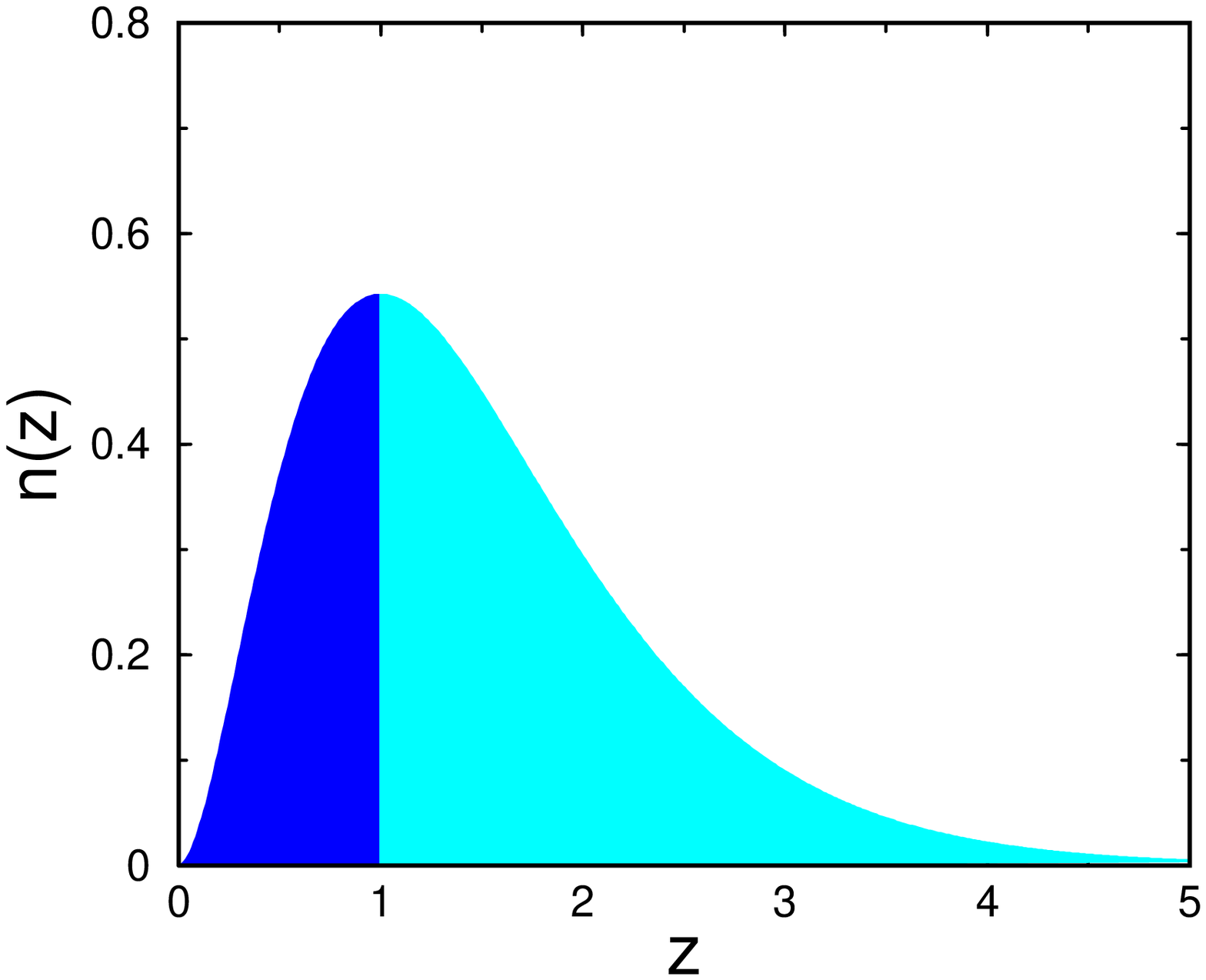, height=2.0in, width=3.0in}
\epsfig{file=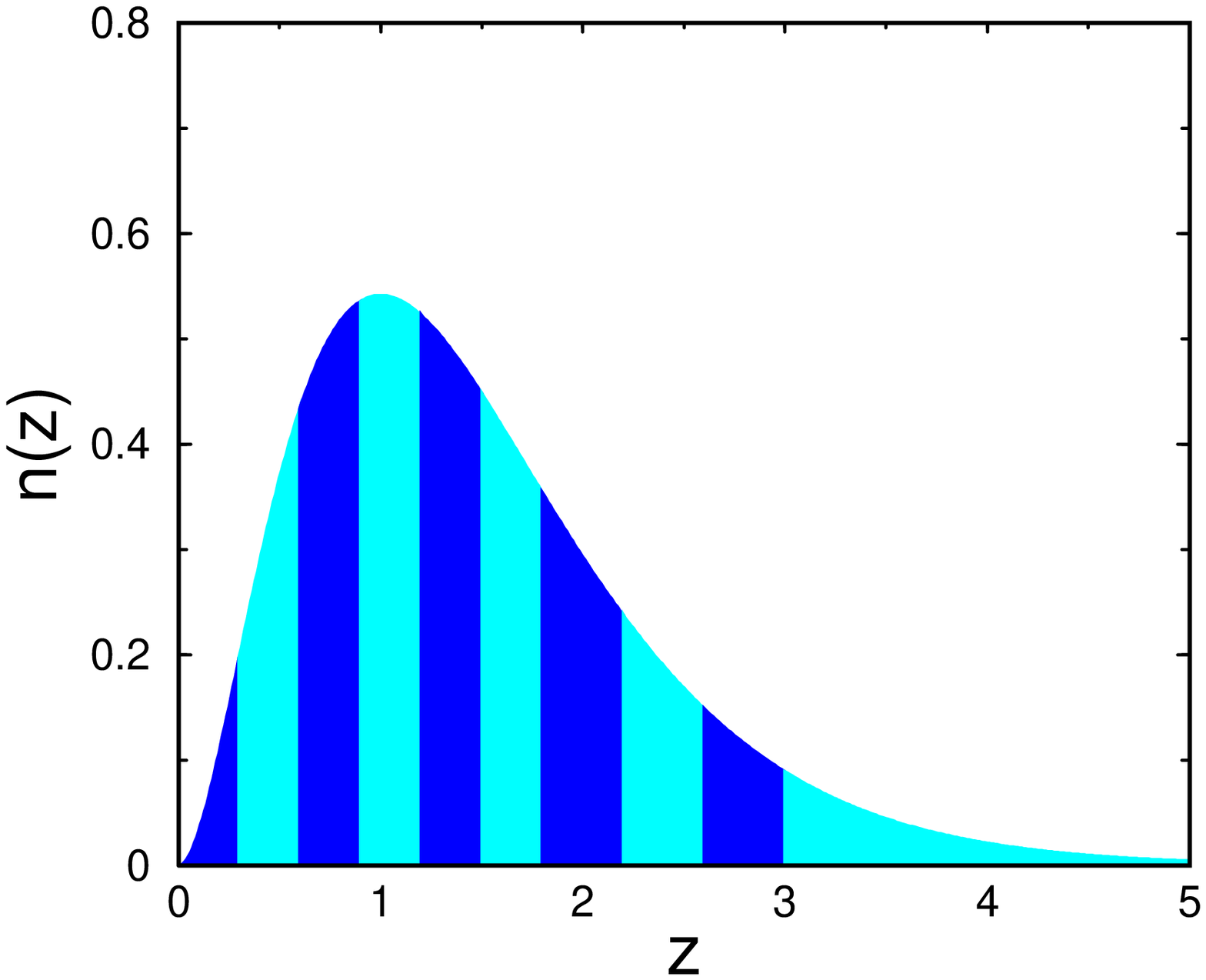, height=2.0in, width=3.0in}
\caption{The divisions of source galaxies in redshift we used in
order to implement the tomography. {\it Top panel:} A simple
division in two redshift bins. {\it Bottom panel:} A division in
10 redshift bins.
\label{fig:tomo_divide} }
\end{center}
\end{figure}

\begin{figure}[htbp]
\begin{center}
\epsfig{file=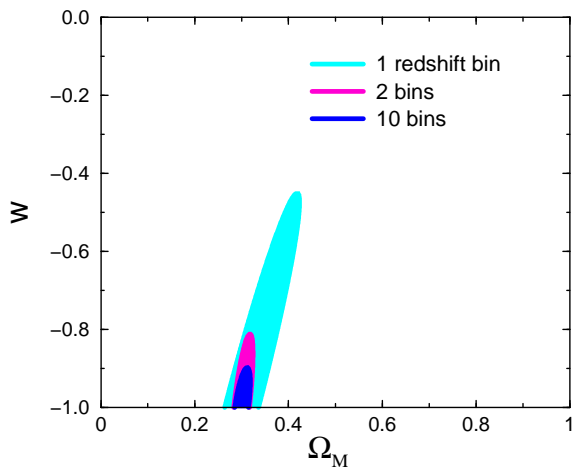, height=2.5in, width=3.0in}
\caption{The improvement in the constraint on $\Omega_M$ and $w$ due
to tomography. The 68\% C.\L.\ constraint regions correspond to 1, 2
and 10 divisions in redshift (largest to smallest ellipse), and are
all computed using the Planck (T) prior.
\label{fig:tomo_constr} }
\end{center}
\end{figure}

Following the formalism of Hu \cite{tomography}, we compute the
parameter constraints when source galaxies are separated in redshift.
Of the several slicings in two bins we tried, the most effective 
division was below and above $z=1.0$
(Fig.~\ref{fig:tomo_divide}, top panel). In this case, the constraints
on $\Omega_X$ and $w$ improve by a factor of 3 and 1.4 respectively,
for a Planck(T) prior (Fig.~\ref{fig:tomo_constr}). For the weaker
COBE+photo-z prior, the improvement is even more significant: a factor
of 5 and 3 improvements on $\Omega_X$ and $w$ respectively.

We also consider an optimistic scenario where galaxies can be
separated in 10 redshift bins (Fig.~\ref{fig:tomo_divide}, bottom
panel).  Whether or not and how accurately something like this can be
done using photometric redshift techniques is presently under
investigation (Eisenstein, Hu and Huterer, in preparation).  The
constraints on $\Omega_M$ (or $\Omega_X$) and $w$ further improve:
$\sigma(\Omega_X)=0.012$ and $\sigma(w)=0.07$  (Fig.~\ref{fig:tomo_constr}).

%

Subdividing the galaxy population in more than two redshift bins
leads to fairly limited improvements in parameter
determination; this is due to high correlations ($\sim$80\%)
between the power spectra in different bins \cite{tomography}.
Nevertheless, tomography clearly adds valuable information on
cosmological parameters and should be pursued with data from
future WL experiments.  In order to accurately assess and
optimize this technique, further study considering realistic
accuracy of photometric redshifts is necessary.  Using simplified
assumptions (in particular, no ``leakage'' of galaxies between
bins), we have shown here that separation of galaxies in redshift
easily leads to a factor of a few improvement in measuring $\Omega_X$
and $w$.
 
We now discuss whether a specific signature of certain dark-energy models
can be detected with WL surveys.

\subsection{Detecting the Dark-Energy Clustering?}\label{QCDM_clustering}

Evolving scalar fields, or quintessence, are a particular class
of candidates for dark energy (e.g. \cite{Wetterich, Coble,
Zlatev}).  One signature of quintessence is that it generally
clusters around and above the Hubble radius scale $H_0^{-1}$.  We
ask: is it possible to detect this clustering in wide-field weak
lensing surveys?

The clustering of quintessence is reflected in the increase in
the transfer function on very large scales. The effect is more
pronounced for larger $\Omega_X$ and larger $w$, and explicit
forms for $T_Q(k, z)$ are given in Refs.~\cite{Ma_QCDM} and
\cite{dark_synergy}; here $T_Q$ is the transfer function that
takes clustering into account.

\begin{figure}[htbp]
\begin{center}
\epsfig{file=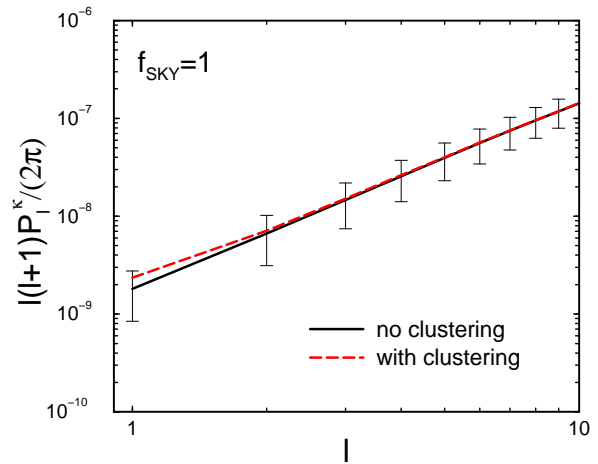, height=2.5in, width=3.0in}
\caption{The effect of clustering of quintessence on the
convergence power spectrum for a fiducial equation-of-state ratio
$w=-1/3$. The error bars correspond to the  cosmic variance for a
full-sky weak lensing survey. Clustering affects the $l=1$
multipole the most, but even there the effect is buried within cosmic
variance.
\label{fig:pk_l.nolimber}}
\end{center}
\end{figure}

Clustering changes the matter power spectrum on large scales,
which in turn alters the convergence power spectrum at lowest
multipoles. In Figure~\ref{fig:pk_l.nolimber} we show an
optimistic scenario\footnotemark[2]
\footnotetext[2]{In a sense that a more positive $w$
leads to more clustering.} with $w=-1/3$ with and without
clustering taken into account. We used exact formulae for the
convergence power spectrum \cite{CMB_lens}, since Limber's
approximation breaks down at lowest multipoles.  Even though the
effect on the matter power spectrum is significant ($T_Q(k,
z)/T_{\Lambda}(k, z)\approx 2.0$ at $k\sim H_0$ in this
case), the convergence power spectrum changes noticeably only at
$l=1$, and even there only by $\sim 30\%$. As this Figure shows,
the effect is buried deeply in the cosmic variance even for a
full-sky WL survey. Therefore, it is unlikely that WL
alone can detect the clustering of quintessence. However,
cross-correlation of WL and other methods (e.g.\ the CMB) may be
more promising; see Refs.~\cite{Kinkhabwala, Peiris}.

\section{Systematics and biases}\label{sec:sys}

\subsection{Observational issues}

It is difficult to overemphasize the importance of controlling
the various systematic errors that generically creep into the WL
observing process. These include shear recovery issues,
anisotropic point-spread function, the quality of seeing, and
instrumental noise (for a nice study of systematic effects, see
Ref.~\cite{sys_study}). There is also the effect of overlapping
galaxies, which is expected to be especially pronounced in very
deep surveys, but might be overcome using the photometric
redshift information (M. Joffre, private communication). Finally,
the observed galaxies might be intrinsically aligned due to
coupling of their angular momenta or a similar mechanism
(\cite{Crittenden, Croft_Metzler, Heavens} and references
therein); this has already been observed \cite{Pen, Brown}. These
effects may masquerade as the signal itself, and make the
extraction of ellipticity correlations very difficult. In our
analysis, we have assumed that these problems will be resolved,
and that the dominant uncertainty will be the cosmic variance on
large scales and Poisson noise on small scales. In that sense,
our results (for a given parameter space, set of priors, and
fiducial survey strategy) may be optimistic. On the other hand,
rapid advances in our understanding of weak lensing techniques,
as well as the prospects of powerful future surveys, indicate
that in a few years we can expect a much better understanding of
the aforementioned problems.

\subsection{Dependence upon nonlinear power spectrum and galaxy
distribution}\label{sec:NLPS}

In addition to observational systematics that need to be controlled,
theoretical prediction for the angular power spectrum of the
convergence is also uncertain. Uncertainties in the nonlinear matter
power spectrum (NLPS) and in the redshift distribution of galaxies are
especially significant, as they are difficult to quantify and were not
included in our analysis.  We now discuss these two ingredients in
more detail.

As can be seen from Figs.~\ref{fig:pk_l} and \ref{fig:matter_ps},
most of our constraints come from nonlinear scales. Therefore,
knowledge of the NLPS is crucial in order to compare experimental
results with theory. However, this quantity is perhaps the most
uncertain ingredient in the prediction for the power spectrum of
the convergence. The NLPS is traditionally obtained by running
N-body simulations for several cosmological models and deriving a
fitting  function to the simulated nonlinear power spectra. For
the models with dark energy we consider, either PD or Ma
QCDM fitting functions can be used. The latter was
calibrated for quintessence models in a flat universe, and tested
at $w=-2/3$, $-1/2$ and $-1/3$ and $\Omega_M=0.4$ and $0.6$.

Even with this solution, the intrinsic uncertainty of 5-15\% in
the NLPS is significant (recall, the transfer and growth
functions are accurate to just a few percent). To illustrate the
importance of knowing the NLPS accurately, let us for the moment
{\it assume} that the true NLPS at $w=-1$ is that given by the
formula of PD. Let us further {\it assume} that, not knowing
this, we adopt the Ma QCDM prescription to compute the
theoretical power spectra.
We now compute the bias in cosmological parameters due to this
``erroneous'' assumption. Let us write the cosmological parameter
values as

\begin{equation}
p_i=\bar p_i + \delta p_i
\end{equation}

\noindent where $\bar p_i$ is the true value, $p_i$ the measured value, and
$\delta p_i$ the bias due to using the ``wrong'' NLPS.  Assuming that
these biases are small, it is easy to show that \cite{Knox}

\begin{equation}
\delta p_i = F_{ij}^{-1} \sum_l
{1\over (\Delta P_l^{\kappa})^2} 
(P_l^{\kappa}-\bar P_l^{\kappa})
{\partial \bar P_l^{\kappa} \over \partial p_j}
\label{eq:bias}
\end{equation}

\noindent where $F_{ij}$ is the ubiquitous Fisher matrix,
$P_l^{\kappa}$ ($\bar P_l^{\kappa}$) is the ``erroneous''
(``true'') power spectrum, and sum over $j$ is implied.  The
results of this exercise are given in Table 2 where we consider
our fiducial survey with Planck (T) prior. The biases in
$\Omega_X$ and $w$ are 2.5 and 4.8 times the 1-$\sigma$
uncertainties in these parameters! Even though these numbers may
not be accurate because the approximation $\delta p\ll \bar p$
necessary to use Eq.~(\ref{eq:bias}) obviously did not hold, one
can still conclude that the biases are very
significant. Therefore, we need a more accurate knowledge of the
NLPS. 

\begin{center}
{TABLE 2\\[4pt] \scshape \rule[-2mm]{0mm}{5mm} Parameter Biases}
\nopagebreak
\begin{tabular}{lccccc}\hline\hline
      & \multicolumn{2}{c}{\rule[-2mm]{0mm}{6mm}Due to ``wrong'' NLPS}  
      &
      & \multicolumn{2}{c}{\rule[-2mm]{0mm}{6mm}Due to ``wrong'' $n(z)$}\\
					\cline{2-3} \cline{5-6}
\rule[-3mm]{0mm}{8mm}$p_i$ & $|{\rm bias}|$  & $|{\rm bias}|/\sigma(p_i)$ &
			   & $|{\rm bias}|$  & $|{\rm bias}|/\sigma(p_i)$ 
						\\\hline\\[-0.15cm]

$\Omega_X$
	&	$0.09$
	&	$2.5$ 
	&
	&	$0.04$
	&	$1.2$ \\[0.1cm]
$w$
	&	$0.92$ 
	&	$4.8$ 
	&
	&	$0.57$
	&	$3.0$ \\[0.1cm]
\label{tab:bias}
\end{tabular}
\end{center}

Fortunately, the NLPS obstacle is surmountable. It is a matter of
running powerful N-body simulations that include dark energy, on
a fine grid in $w$ (and $m_\nu$ and other parameters, if
necessary).  Because we are only interested in the matter power
spectrum (not galaxy power spectrum, which includes bias), N-body
simulations can in principle give the NLPS to a very  high
accuracy. Once this is achieved, weak lensing will regain much of
its power to probe dark energy.

Another quantity that may not be known to an extremely high
accuracy (although we assumed so) is the redshift distribution of
source galaxies $n(z)$. Indeed, current photometric redshift
techniques can determine redshifts to an accuracy of $\sim 0.1$,
depending on the redshift (e.g., \cite{Hogg}), which leaves
room for error, both statistical and systematic. To include the
uncertainty in $n(z)$, some authors (e.g., \cite{Jain_Seljak,
Bernardeau, Hu_Tegmark}) parameterized the
redshift distribution by one parameter only. However, the
realistic uncertainty in $n(z)$ is much more difficult to
quantify. To assess the effect of an uncertainty in the redshift
distribution, we assume that the true distribution is given by
Eq.~(\ref{eq:prob_tys}) with $z_0=0.5$, while we ``erroneously''
assume the same form with $z_0=0.55$ (recall, $n(z)$ peaks at
$z=2z_0$). The biases  in $\Omega_X$ and $w$ are given in Table
2, and are 1.2 and 3.0 times the unbiased  1-$\sigma$ uncertainties in
these parameters, respectively.  Just as in the case of the NLPS,
we conclude that accurate knowledge of the redshift distribution
of galaxies will be crucial if weak lensing is to achieve its
full potential.

\subsection{Power spectrum covariance}\label{sec:covar}

Yet another important issue that we ignored so far is covariance
of the convergence power spectrum. The shear (or convergence)
field is expected to be non-Gaussian due to nonlinear
gravitational processes. Therefore, measurements of
$P_l^{\kappa}$ are generally going to be correlated, implying a
non-zero four-point function (or its Fourier analogue, the
trispectrum). The covariance will be especially pronounced at
high multipoles. For a survey down to a limiting magnitude of
$R\sim 25$, the effect of power spectrum covariance appears to be
small: Cooray \& Hu \cite{Hu_covar} have used the dark-matter
halo approach to compute the power spectrum as well as the
trispectrum, and found that the non-Gaussianity increases errors
on cosmological parameters by about 15\%. Although this effect is
small enough to be ignored with current datasets, it will be
important to take it into account when interpreting results from
upcoming deep surveys because the covariance on small scales is
likely to significantly degrade the cosmological
constraints. Restricting our analysis (with COBE+photo-z prior)
to multipoles $l\leq 3000$ degrades the constraints on $\Omega_X$
and $w$ by a factor of 5. Clearly, information from small scales is
important, and it will be necessary to carefully assess the
impact of power spectrum covariance for deep WL surveys.

\section{Three-point statistics and dark energy}\label{sec:3pt}

We now turn to three-point statistics of the weak lensing
convergence. Unlike the CMB temperature fluctuations which may or
may not be Gaussian, weak lensing convergence almost certainly
does not obey Gaussian statistics. In this Section, we illustrate
the dependence of the bispectrum and skewness of the convergence
on dark energy, and show that they present a promising avenue
that can lead to the dark component.

\subsection{Preliminaries}

The bispectrum of the convergence $B_{l_1 l_2 l_3}^\kappa $ is
defined through the three-point correlation function of the
convergence in multipole space 

\begin{equation}
\left< \kappa_{l_1 m_1} \kappa_{l_2 m_2} \kappa_{l_3 m_3} \right> 
= \wjm B_{l_1 l_2 l_3}^\kappa \,
\end{equation}

\noindent and can further be written as

\begin{eqnarray}
B_{l_1 l_2 l_3}^\kappa &=& \sqrt{(2l_1+1)(2l_2+1)(2l_3+1)  \over 4\pi} \wj
        \nonumber\\
&\times& \left[ \int d\chi {[W(\chi)]^3 \over r(\chi)^4}  B\left({l_1 \over
        r(\chi)},{l_2 \over r(\chi)},{l_3\over r(\chi)},\chi\right)\right] \, 
\label{eq:bispectrum}
\end{eqnarray}

The bispectrum is defined only if the following relations are
satisfied: $|l_j-l_k|\leq l_i \leq |l_j+l_k|$ for $\{i, j, k\}\in
\{1, 2, 3\}$ and $l_1+l_2+l_3$ is even. The term in parentheses
is the Wigner 3j symbol, which is closely related to
Clebsch-Gordan coefficients from quantum mechanics (for its
properties, see Refs.~\cite{Varshalovich,  Hu_reion}).
$W(\chi)$ is the weight function defined in
Sec.~\ref{sec:prelim}. To compute the bispectrum of the
convergence, therefore, we need to supply the matter bispectrum
$B(k_1, k_2, k_3,  z)$. The latter quantity can be calculated in
linear theory (that is, on large scales), but, just as in the
case of the matter power spectrum, it needs to be calibrated from
N-body simulations on nonlinear scales. Here we adopt the fitting
formulae of Scoccimarro \& Couchman (\cite{SC}; heretofore SC) which
are based on numerical simulations due to VIRGO collaboration
\cite{VIRGO}. The matter bispectrum is defined only for
closed-triangle configurations ($\vec k_1+ \vec k_2+ \vec k_3=0$)
and is given by

\begin{equation}
B(\vec k_1, \vec k_2, \vec k_3)=
2\,F_2(\vec k_1, \vec k_2) P(k_1) P(k_2)+{\rm cycl. }
\end{equation}

\noindent  where $P(k)$ is the matter power spectrum and

\begin{eqnarray}
F_2(\vec k_1, \vec k_2)&=& {5\over 7} a(n, k_1)\,a(n, k_2) +\nonumber\\ 
	&&{1\over 2} {\vec k_1 \cdot\vec k_2 \over k_1\,k_2}
	\left ({k_1\over k_2}+{k_2\over k_1}\right )b(n,
	k_1)\,b(n, k_2) +
	\nonumber\\
&&{2\over 7} \left ({\vec k_1 \cdot\vec k_2 \over k_1\,k_2}\right )^2
	c(n, k_1)\,c(n, k_2)
\end{eqnarray}

\noindent $n\equiv d\ln P/d\ln k$, and functions $a$, $b$ and $c$ are
given in SC. Although not explicitly tested on models involving
dark energy, the fitting formula depends on cosmology only
through the matter power spectrum; this weak dependence on
cosmology is also borne out in high-order perturbation theory
\cite{SCFFHM}. Therefore, we decide to use the SC
formula to illustrate the dependence of three-point statistics on
dark energy.
 
\begin{figure}[htbp]
\begin{center}
\epsfig{file=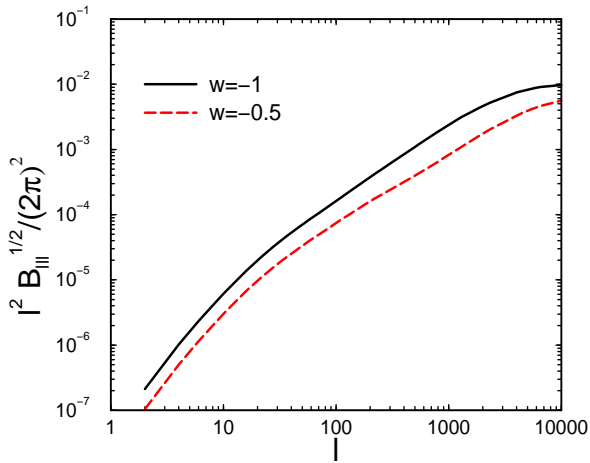, height=2.5in, width=3.0in}
\caption{The quantity $l^2 B^{1/2}_{lll}/(2\pi)$, involving
equilateral-triangle configurations of the bispectrum in multipole
space. We use this quantity to illustrate how the bispectrum depends
on dark energy. The variance in $B_{lll}$ is
roughly two orders of magnitude larger than the signal. }
\label{fig:bispec}
\end{center}
\end{figure}

In Fig.~\ref{fig:bispec} we show the quantity $l^2
\sqrt{B_{lll}^{\kappa}}/(2\pi)$ \cite{Hu_bispec} for $w=-1$
and $w=-0.5$; here $B_{lll}^{\kappa}$ is the equilateral triangle
configuration of the bispectrum\footnotemark[2]
\footnotetext[2]{We set
$m_{\nu}=0$ in this Section.}. Since roughly $ B\propto P^2$ and
$B$ has little other dependence on dark energy, we expect
that $l^2 \sqrt{B_{lll}^{\kappa}}/(2\pi)$ varies with $w$
similarly as $P$ --- and this is correct (compare
Figs.~\ref{fig:pk_l} and \ref{fig:bispec}). Therefore, the
bispectrum appears to be an excellent probe of dark
energy. Things are complicated, however, by the large cosmic
variance of a bispectrum. Although computing variance of $B$
involves a daunting task of evaluating the six-point correlation
function of the convergence, this quantity can be computed under
an assumption of small departures from Gaussianity \cite{Luo, 
Gangui_Martin}. For the equilateral triangle
configuration of the bispectrum we show, this estimate indicates
that the cosmic variance is about two orders of magnitude larger
than the bispectrum signal itself, roughly independently of
$l$. Therefore, it is unlikely that a single configuration of the
bispectrum can be used to probe dark energy. However, one
should be able to find an optimal combination of configurations in
order to maximize the amount of information. We relegate this
problem to future work.

Next we discuss the dependence of skewness on dark
energy. Skewness is defined as 

\begin{equation}
S_3(\theta) =
\frac{\left<\kappa^3(\theta)\right>}{\left<\kappa^2(\theta)\right>^2}
\, 
\end{equation}

\noindent where

\begin{eqnarray}
\left< \kappa^2(\theta) \right> &=&
{1 \over 4\pi} \sum_l (2l+1) P_l^\kappa \W_l^2(\theta)\\
\left< \kappa^3(\theta) \right> &=&
                {1 \over 4\pi} \sum_{l_1 l_2 l_3}
                \sqrt{(2l_1+1)(2l_2+1)(2l_3+1) \over 4\pi} \nonumber\\
                &&\times \wj B_{l_1 l_2 l_3}^\kappa
                \W_{l_1}(\theta)\W_{l_2}(\theta)\W_{l_3}(\theta)
                \,. \nonumber \\
\end{eqnarray}

\noindent are the second and third moments of the map smoothed
over some angle theta, and $\W_l(\theta)$ is the Fourier
transform of the top-hat function:
$\W_l(\theta)=2J_1(l\theta)/(l\theta)$. Skewness effectively
combines many different bispectrum configurations, and its
variance should be much smaller than that of $B_{l_1 l_2 l_3}$.  Its
disadvantage is that measurements on different scales are
correlated.

\begin{figure}[htbp]
\begin{center}
\epsfig{file=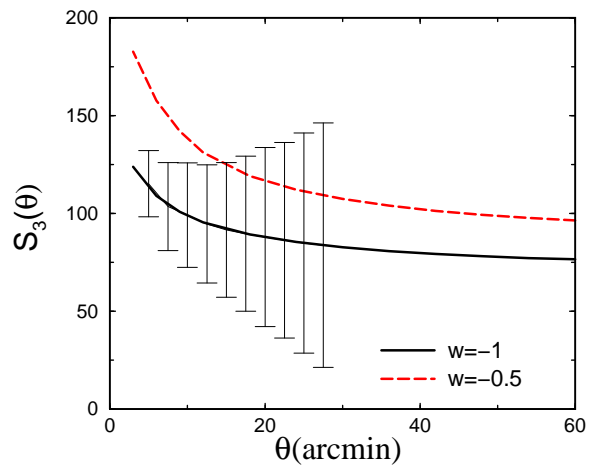, height=2.5in, width=3.0in}
\caption{Skewness of the convergence for two values of $w$.
Error bars are from simulations by White \& Hu \cite{White_Hu} on
scales they explore and for a field of 36 sq.\ deg.}
\label{fig:skew}
\end{center}
\end{figure}

Fig.~\ref{fig:skew} shows skewness for two values of $w$.
Roughly speaking, $S_3\propto B^{\kappa}/P^2$, and although
$B^{\kappa}$ and $P^2$ both decrease with increasing $w$, the
$P^2$ term prevails --- hence the scaling of $S_3$ with $w$. The
error bars shown are those from White \& Hu \cite{White_Hu} for
their WL simulations corresponding to the $\Lambda$CDM model, and
for a field of 36 sq.\ deg. Although the dependence of skewness
on dark energy is significant, there are several obstacles.  As
in the case of the matter power spectrum, the fitting formula for
the bispectrum is accurate only to about 15\% (rms deviation) for
$\Lambda$CDM models and not yet calibrated for dark energy
models.  More seriously, the measurements of skewness are likely
to be highly correlated --- in fact, van Waerbeke et al.\
\cite{vW_Scocc} find that correlation between skewness
measurements (for the top-hat filter we use) is close to 100\%.

In conclusion, our preliminary analysis indicates that the
three-point statistics of the weak lensing convergence are
sensitive to the presence of dark energy, mainly through the
dependence of the matter power spectrum. More work is needed,
however, in order for the three-point statistic to become an
effective probe of the missing component. This will include
sharpening the predictions for the three-point function in the
nonlinear regime, and finding optimal configurations of
the bispectrum to probe dark energy.

\section{Discussion and Conclusions}\label{sec:concl}

Recent results coming from type Ia supernovae, CMB, and
large-scale structure surveys make a strong case for the
existence of dark energy. It is therefore important to
explore how upcoming and future surveys can be used to probe this
component.  In this work, we explore the power of weak
gravitational lensing to probe dark energy via its
measurements of the power spectrum of the convergence.

Dark energy modifies the convergence power spectrum by altering
the distance-redshift relation, as well as the matter power
spectrum. The dependence on dark energy is therefore somewhat
indirect, and cannot be easily disentangled from the effect of
other parameters ($\Omega_M h^2$, $\Omega_B h^2$, $n$, $m_\nu$).
Because of this, one would not expect WL to be an efficient way
to probe dark energy.  Nevertheless, we find that with the
proposed very wide and deep surveys, WL can be an important
probe of dark energy, on a par with SNe Ia and number counts. We
consider a generic future survey covering 1000 sq.\ deg.\ down to
a limited magnitude of $R=27$, cosmic-variance limited on large
scales and Poisson-noise limited on small scales.  With
photometric redshift information and constraints on other
parameters that would be expected from the Planck experiment with
temperature information only, we find that such a survey is able
to constrain $\Omega_X$ and $w$ to between a few percent and a
few tens of percent, depending on the fiducial model and a chosen
set of priors. The constraints are in general stronger for wider
and deeper surveys, and depend on the fiducial model (e.g., the
neutrino mass).  Accurate knowledge of the redshift distribution
of source galaxies will be crucial; we find that an error of only
$0.05$ in the peak of the redshift distribution can bias the
results.

There are important caveats to this result, however. Most
information from WL comes from nonlinear scales, where the
evolution of density perturbations is difficult to track
analytically and understood mostly through N-body simulations
(restricting the analysis of $P_l^{\kappa}$ only to linear scales
with $l\lesssim 100$ would lead to extremely weak constraints on
cosmological parameters due to cosmic variance). The
nonlinearities potentially lead to at least two sources of
systematic error. First, the power spectrum measurements
$P_l^{\kappa}$ are likely to be strongly correlated at multipoles of
several thousand and higher. This is especially true for planned
deep surveys (down to a limited magnitude of $R\sim 27$ or
higher), and these correlations will likely degrade the
constraints on $\Omega_X$ and $w$. 

Second, although the nonlinear power spectrum has been calibrated
quite accurately for $\Lambda$CDM models, most notably through
the PD formula, it remains poorly explored for models with
general equation of state $w$, massive neutrinos, and significant
baryon density. We explicitly showed that the lack of knowledge of
the dependence of the nonlinear power spectrum on $w$ can easily
bias the constraints on $\Omega_X$ and $w$. Therefore, a better
understanding and calibration of the NLPS is absolutely crucial
in order to use WL as a tool of precision cosmology.  

This problem can be turned around, however. One could use the
weak lensing measurements (combined, perhaps, with information
from CMB measurements, SNe Ia, and other probes) in order to
{\it constrain} the nonlinear power spectrum.  This constraint
could be very interesting, given the strong dependence of the
NLPS on cosmological parameters.

Predictions for the three-point statistics of WL are quite
uncertain at present, especially for models involving dark
energy.  This does not mean they will not become effective probes
of the missing component in the future. We estimate the
equilateral bispectrum configuration, as well as skewness, for
two values of $w$ and show that dependence on $w$ is
significant. Although these two quantities are plagued by large
cosmic variance and highly correlated noise respectively, by
clever choice of bispectrum configurations one might be able to
increase the signal-to-noise ratio and extract useful information
on dark energy.

There are other ways to use weak lensing as a probe of cosmology which
we did not discuss. For example, one could use WL to identify clusters
of galaxies at redshifts $0<z\lesssim 3$ (M. Joffre et al., in
preparation). Comparing the measured number density of clusters to the
prediction given by the formalism of Press \& Schechter \cite{PS} gives
constraints on cosmology. Another idea is to measure the angular power
spectrum of clusters (detected through WL) at different redshifts
\cite{cluster_PS}; this gives a direct measure of the angular
diameter distance as a function of redshift. The advantage of this
approach is that only the {\it linear} matter power spectrum is
required; furthermore, the mass function and profiles of clusters need
not be known. These two methods will provide constraints complementary
to those from the galaxy shear.

Weak gravitational lensing is likely to provide a wealth of
information not only on the matter distribution in the universe,
but also on the amount and nature of dark energy. We have
considered the basic program of measuring the convergence power
spectrum, and found that very wide and deep surveys could provide
information complementary and comparable to that from other
cosmological probes. Other statistics (various bispectrum
configurations, cross-correlation of WL and the CMB, etc.) are
likely to further increase the power of weak lensing and make it
an important probe of dark energy.

\bigskip

\begin{acknowledgments}
The author would like to thank Asantha Cooray, Vanja Duki\'{c},
Joshua Frieman, Gil Holder, Mike Joffre, Michael Turner, and
especially Wayne Hu for many useful discussions.  This work
represents partial fulfillment of the requirements for the
Ph.D. in Physics at the University of Chicago, and was supported
by the DoE.
\end{acknowledgments}


\begin{thebibliography}{99}

\bibitem{Riess}
	A. Riess {\it et al.}, Astron. J., {\bf 116}, 1009 (1998)

\bibitem{Perlmutter}
	S. Perlmutter {et al.},  Astrophys. J. {\bf 517}, 565 (1999)


\bibitem{reconstr}
	D. Huterer and M. S. Turner, Phys. Rev. D {\bf 60}, 081301 (1999)

\bibitem{Weller}
	J. Weller and A. Albrecht, report DAMTP-2001-53 (astro-ph/0106079)

\bibitem{Maor}
	I. Maor, R. Brustein, and P. J. Steinhardt, Phys. Rev. Lett. {\bf
	86}, 6 (2001)

\bibitem{Goliath}
	M. Goliath, R.  Amanullah, P. Astier, A. Goobar, A. and
	R. Pain, astro-ph/0104009

\bibitem{Turner_scripta}
	M. S. Turner, Physica Scripta {\bf T85}, 210 (2000)

\bibitem{optimal}
	D. Huterer and M. S. Turner, Phys. Rev. D in press
	(astro-ph/0012510) 

\bibitem{Holder}
	Z. Haiman, J. J. Mohr, and G. P. Holder,
	Astrophys. J. {\bf 553}, 545 (2001)

\bibitem{Davis}
	J. Newman and M. Davis, Astrophys J. {\bf 534 },
	L11 (2000)

\bibitem{early_WL}
	 J. A. Tyson, F. Valdes, J. F. Jarvis and A. P. Jr. Mills,
	 Astrophys. J., {\bf 281}, L59 (1984)

\bibitem{Jordi} 
	J. Miralda Escud\'{e}, Astrophys. J., {\bf 380}, 1 (1991)

\bibitem{Blanford} 
	R. Blanford, A. Saust, T. Brainerd and J. Villumsen, 
	Mon. Not. R. Astron. Soc., {\bf 251}, 600 (1991)

\bibitem{Kaiser_92}
	N. Kaiser, Astrophys. J., {\bf 388}, 272 (1992)

\bibitem{Kaiser_98}
	N. Kaiser, Astrophys. J., {\bf 498}, 26 (1998)

\bibitem{Jain_Seljak}
	B. Jain and U. Seljak, Astrophys. J., {\bf 484}, 560 (1997)

\bibitem{Kamionkowski} 
	M. Kamionkowski, A. Babul, C. M. Cress, and A. Refregier,
	Mon. Not. R. Astron. Soc., {\bf 301}, 1064 (1998)

\bibitem{Kaiser_Squires}
	N. Kaiser and G. Squires, Astrophys. J., {\bf  404}, 441 (1993)

\bibitem{Mellier}
	Y. Mellier,  Proceedings of the NATO Advanced Study Institute
                     on Theoretical and Observational Cosmology, ed.\
                     Marc Lachieze-Rey; Kluwer Academic, 1999.;
                     astro-ph/9901116 

\bibitem{Wittman} 
	D. Wittman, J. A. Tyson, D. Kirkman, I. Dell'Antonio and 
	G. Bernstein, Nature, {\bf 405}, 143 (2000)

\bibitem{Bacon} 
	D. J. Bacon, A. R. Refregier and R. S. Ellis,
	Mon. Not. R. Astron. Soc., {\bf 318}, 625 (2000)

\bibitem{vW}
	L. Van Waerbeke {\it et al.}, A\&A {\bf 358}, 30 (2000)

\bibitem{Kaiser_00} 
	N. Kaiser, G. Wilson and G. A. Luppino, astro-ph/0003338

\bibitem{Turner_White}
	M. S. Turner and M. White, Phys. Rev. D, {\bf 56}, R4439 (1997)

\bibitem{Bernardeau} 	
	F. Bernardeau, L. Van Waerbeke and Y. Mellier, A\&A, {\bf
	322}, 1 (1997)

\bibitem{Hu_Tegmark}
	W. Hu and M. Tegmark, Astrophys. J., {\bf 514}, L65 (1999)

\bibitem{boom}
	C. B. Netterfield {\it et al.} astro-ph/0104460

\bibitem{DASI}
	C. Pryke {\it et al.}, astro-ph/0104490

\bibitem{max}
	R. Stompor {\it et al.}, astro-ph/0105062

\bibitem{Wang}
	X. Wang, M. Tegmark and M. Zaldarriaga, astro-ph/0105091

\bibitem{Bartelmann}
	M. Bartelmann and P. Schneider, Physics Reports, {\bf
	340}, 291 (2001)

\bibitem{Hu_White} 
	W. Hu and M. White, Astrophys. J., {\bf 519}, L9 (2000)

\bibitem{Broadhurst}
	T. Broadhurst, astro-ph/9505010

\bibitem{Bennett}
	C. L. Bennett {\it et al.}, Astrophys. J., {\bf 464}, L1 (1996)

\bibitem{Ma_QCDM}
	C.-P. Ma, R. R. Caldwell, P. Bode and L. Wang, Astrophys. J.,
	{\bf 521}, L1 (1999) (Ma QCDM)

\bibitem{Bunn_White}
	E. F. Bunn and M. White, Astrophys. J., {\bf 480}, 6 (1997)

\bibitem{Liddle_Lyth}
	A. Liddle and D. Lyth, "Cosmological inflation
	and large-scale structure", Cambridge University Press (2000)

\bibitem{Hu_transfer}
	W. Hu and D. J. Eisenstein, Astrophys. J., {\bf 498}, 497 (1998)

\bibitem{Carroll}
	S. M. Carroll, W. H. Press and E. L. Turner, ARA\&A
	{\bf 30}, 499 (1992)

\bibitem{Peebles}
	P. J. E. Peebles, "Principles of Physical Cosmology",
	Princeton University Press (1993)

\bibitem{Hamilton}	
	A. J. S. Hamilton, A. Matthews, P. Kumar and E. Lu,
	Astrophys. J., {\bf 374}, L1 (1991)

\bibitem{PD} 
	J. A. Peacock and S. J. Dodds, Mon. Not. R. Astron. Soc.,
	{\bf 267}, 1020 (1994) (PD)

\bibitem{Ma_LCDM}
	C.-P. Ma, Astrophys. J., {\bf 508}, L5 (1998) (Ma LCDM)

\bibitem{Fisher}	
	M. Tegmark, A. N. Taylor, and A. F. Heavens,
	Astrophys. J. {\bf 480}, 22 (1997)

\bibitem{Zaldarriaga_Seljak}
	M. Zaldarriaga and U. Seljak, Phys. Rev. D., {\bf 55}, 1830 (1998)

\bibitem{Eis_Hu_Teg}
	D. Eisenstein, W. Hu, M. Tegmark, Astrophys. J. {\bf
	518}, 2 (1999)

\bibitem{SN_Fisher}
	M. Tegmark, D. J. Eisenstein, W. Hu and R. Kron,
	astro-ph/9805117 

\bibitem{Holder_01}
	G. P. Holder, Z. Haiman and J. J. Mohr, ApJL, in press
	Astrophys. J., {\bf 560}, L111 (2001)

\bibitem{Metcalfe} 
	N. Metcalfe, T. Shanks, A. Campos, H. J. McCracken and
	R. Fong, Mon. Not. R. Astron. Soc., {\bf 323}, 795 (2001)

\bibitem{Hogg}
	D. W. Hogg {\it et al.},AJ, {\bf 115}, 1418 (1998)

\bibitem{Ly_alpha}
	R. Croft, W. Hu and R.  Dav\'{e}, 
	Phys. Rev. Lett., {\bf 83}, 1092 (1999)

\bibitem{tomography}
	W. Hu, Astrophys. J., {\bf 522}, L21 (1999)

\bibitem{Wetterich}
	C. Wetterich, Nucl. Phys. B {\bf 302}, 668 (1988)

\bibitem{Coble}
	K. Coble, S. Dodelson, and J. Frieman, Phys. Rev. D {\bf
	55}, 1851 (1996)

\bibitem{Zlatev}
	I. Zlatev, L. Wang and P. J. Steinhardt,
	Phys. Rev. Lett. {\bf 82}, 896 (1999)

\bibitem{dark_synergy}
	W. Hu, Phys. Rev. D, submitted (astro-ph/0108090)
 
\bibitem{CMB_lens}
	W. Hu, Phys. Rev. D, {\bf 62}, 043007 (2000)

\bibitem{Kinkhabwala}
	A. Kinkhabwala and M. Kamionkowski,
	Phys. Rev. Lett., {\bf 82}, 4172 (1999)

\bibitem{Peiris}
	H. V. Peiris and D. N. Spergel, 2000, Astrophys. J., {\bf
	540}, 605 (2000)
 
\bibitem{sys_study}
	D. J. Bacon, A. Refregier, D. Clowe and R. S. Ellis, 	
	Mon. Not. R. Astron. Soc., {\bf 325}, 1065 (2001)	

\bibitem{Crittenden}
	R. G. Crittenden, P. Natarajan, U.-L. Pen and T. Theuns,
	Astrophys. J., {\bf 559}, 552 (2001)

\bibitem{Croft_Metzler}
	R. Croft and C. Metzler, Astrophys. J., {\bf 545}, 561 (2000)

\bibitem{Heavens}
	A. Heavens, A. Refregier and C. Heymans,
	Mon. Not. R. Astron. Soc., {\bf 319}, 649 (2000)

\bibitem{Pen}
	U.-L. Pen, J. Lee and U. Seljak, Astrophys. J., {\bf
	543}, L107 (2000)

\bibitem{Brown} 
	M. L. Brown, A. N. Taylor, N. C. Hambly and S. Dye,
	astro-ph/0009499

\bibitem{Knox}
	L. Knox, R. Scoccimarro and S. Dodelson, 
	Phys. Rev. Lett., {\bf 81}, 2004 (1998)

\bibitem {Hu_covar} 
	A. Cooray and W. Hu, Astrophys. J., {\bf 554}, 56 (2001) 

\bibitem {Varshalovich} 
	D. A. Varshalovich, A. N. Moskalev and V. K. Kheronskii,
	``Quantum THeory of Angular Momentum'', World Scientific (1998)

\bibitem {Hu_reion} 
	A. Cooray and W. Hu, Astrophys. J., {\bf 534}, 533 (2000)

\bibitem{SC}		
	R. Scoccimarro and H. Couuchman,
	Mon. Not. R. Astron. Soc., {\bf 325}, 1312 (2001)	

\bibitem{VIRGO}
	A. R. Jenkins {\it et al.}, Astrophys. J., {\bf 503}, 37 (1998)

\bibitem {vW_Scocc}		
	L. Van Waerbeke, T. Hamana, R. Scoccimarro, S. Colombi and
	F. Berardeau, Mon. Not. R. Astron. Soc., {\bf  322}, 918 (2001)

\bibitem {Hu_bispec} 
	A. Cooray and W. Hu, Astrophys. J., {\bf 548}, 7 (2001)

\bibitem{SCFFHM}		
	R. Scoccimarro, S. Colombi, J. N. Fry, J. A. Frieman,
	E. Hivon and A. Melott, Astrophys. J., {\bf 496}, 586 (1998)

\bibitem{Luo}
	X. Luo, Astrophys. J., { \bf 427} L71 (1994)

\bibitem {Gangui_Martin}
	A. Gangui and J. Martin, Mon. Not. R. Astron. Soc., {\bf
	313}, 323 (2000)

\bibitem {White_Hu}
	M. White and W. Hu, Astrophys. J., {\bf 537}, 1 (2000)

\bibitem {PS}
	W. H. Press and P. L. Schechter, Astrophys. J.,
	{\bf 187}, 425 (1974)

\bibitem {cluster_PS} 
	A. Cooray, W. Hu, D. Huterer and M. Joffre,
	Astrophys. J., { \bf 557} L7 (2001)


\end{thebibliography}
\end{document}